\documentclass[a4paper, 11pt]{article}

\usepackage[makeroom]{cancel}
\usepackage{color}
\usepackage{graphicx}
\graphicspath{{imgs/}{./}}
\usepackage{amsmath}
\usepackage{amssymb}
\usepackage{xspace}
\usepackage[numbers,compress]{natbib}
\usepackage[hyperfootnotes=false]{hyperref}
\usepackage{tcolorbox}
\usepackage{subcaption}
\usepackage{caption}
\usepackage{float}

\newlength{\fighskip} \fighskip=2pt
\newlength{\figvskip} \figvskip=3pt

\newcommand*{\figbox}[2]{{
\def\figscale{#1}
\def\arraystretch{0.8}
\arraycolsep=0pt
\begin{array}{c}
\vbox{\vskip\figscale\figvskip
\hbox{\hskip\figscale\fighskip
\includegraphics[scale=\figscale]{#2}}}
\end{array}}}

\usepackage{mciteplus}
\usepackage{dcolumn}
\usepackage{bm}
\usepackage{verbatim}
\usepackage{amscd}
\usepackage{amsfonts}
\usepackage{setspace}
\usepackage{amsthm}
\usepackage{enumerate}
\usepackage{mathtools}
\usepackage[normalem]{ulem}

\theoremstyle{plain}
\newtheorem{theorem}{Theorem}
\theoremstyle{plain}
\newtheorem{lemma}{Lemma}

\theoremstyle{plain}

\theoremstyle{plain}
\newtheorem{proposition}{Proposition}
\theoremstyle{remark}

\theoremstyle{conjecture}

\theoremstyle{observation}

\theoremstyle{definition}

\theoremstyle{corollary}

\theoremstyle{definition}

\theoremstyle{definition}

\theoremstyle{result}
\newtheorem{definition}{Definition}
\theoremstyle{assumption}

\theoremstyle{definition}

\theoremstyle{problemma}

\theoremstyle{fact}

\newcommand{\BY}[1]{{\textcolor{purple}{BY: #1}}}

\begin{document}
\title{\makebox[\linewidth][r]{\normalfont\normalsize KUNS-3120}\\[1em]\bf A No-Go Theorem for Order-Two Clifford Electric-Magnetic Duality}
\author{
Shunta Takahashi$^{1,3,5}$, Zhi Li$^{2}$, and Beni Yoshida$^{1,3,4,6}$ \\ [1em]
{\em \small $^1$Perimeter Institute for Theoretical Physics, Waterloo, Ontario N2L 2Y5, Canada} \\
{\em \small $^2$IBM Quantum, T.J. Watson Research Center, Yorktown Heights, NY 10598, USA} \\
{\em \small $^3$Department of Physics and Astronomy, University of Waterloo, Waterloo, Ontario N2L 3G1, Canada} \\
{\em \small $^4$Department of Applied Mathematics, University of Waterloo, Waterloo, Ontario N2L 3G1, Canada} \\
{\em \small $^5$Department of Physics, Kyoto University, Kyoto 606-8502, Japan}\\
{\em \small $^6$Fundamental Quantum Science Program (FQSP), RIKEN, Wako 351-0198, Japan}
}
\date{}

\maketitle

\begin{abstract}

Electromagnetic duality in the toric code has order two at the level of anyon types and acts as a Hadamard-type logical transformation on the encoded quantum information. 
Here, we ask whether it can likewise be realized microscopically as an order-two Clifford operation.
For the $\mathbb{Z}_2$ toric code, we prove that any locality-preserving Clifford realization of electric-magnetic exchange cannot have order two.
Our proof is fully general and requires neither translation symmetry with restricted families of system sizes nor a fixed pairing between vertex and plaquette stabilizers. 
Geometrically, the proof locally emulates the unavoidable crossing between electric and magnetic strings, similar to a cross-cap in a non-orientable manifold. 
Extending the problem to the $\mathbb{Z}_N$ toric code, we find that an order-two Clifford realization exists for odd $N$, whereas it is impossible for even $N$.
Our result shows that the order-two exchange of closed-string operators cannot be extended to an order-two exchange of open-string operators.

\end{abstract}

\clearpage

\tableofcontents

\section{Introduction}

Electromagnetic duality is one of the most fundamental examples of duality in physics~\cite{DeserTeitelboim1976, Witten1995}. 
Already in source-free Maxwell theory, the equations are invariant under duality rotations mixing electric and magnetic fields. 
Closely related electric-magnetic dualities appear throughout gauge theory and quantum many-body physics~\cite{Wegner1971, Kogut1979}. 
In many-body systems on a lattice, however, the gauge fields themselves need not be elementary microscopic degrees of freedom.
Instead, they may emerge from collective patterns of quantum entanglement among the underlying local degrees of freedom~\cite{LevinWen2005}.
From this perspective, even the seemingly simple exchange of electric and magnetic fields raises a nontrivial question. 
If the electric and magnetic variables are emergent, what microscopic operation realizes their exchange?

The toric code provides perhaps the simplest setting in which to study this question~\cite{Kitaev2003, DennisKitaevLandahlPreskill2002}.
It realizes an emergent $\mathbb{Z}_2$ gauge theory from microscopic qubits, with electric and magnetic excitations appearing as endpoints of $Z$- and $X$-string operators, respectively.
At the level of the emergent anyon theory, electromagnetic duality is simply the exchange
\begin{align}
e \longleftrightarrow m,
\end{align}
which has order two.
Microscopically, however, this exchange must be implemented by some unitary acting on the underlying qubits.
It is then natural to ask whether such microscopic transformations can obey the same group law,
\begin{align}
U_{em}^2 \propto I.
\end{align}

This microscopic realization is also directly relevant to quantum information processing.
The toric code is one of the simplest and most influential models of quantum error correction, where quantum information is encoded in a degenerate ground-state subspace and protected against local noise and decoherence by many-body entanglement.
At the same time, its Hamiltonian terms, electric and magnetic string operators, and logical Pauli operators are all represented by Pauli operators.
This makes Clifford operations~\cite{Gottesman1997} a natural framework for describing both the microscopic structure of the code and the logical actions induced by anyon permutations, since Clifford unitaries preserve the Pauli group under conjugation~\cite{Bombin2010, Bravyi:2012rnv, Yoshida2015ColorCode}.

Since the electric-magnetic exchange has order two at the level of anyons and acts as a Hadamard-type logical operation, one might therefore expect an order-two microscopic Clifford realization satisfying
\begin{equation}
U_{em}^2 \propto I \ ? \qquad U_{em}\in \text{Clifford}
\end{equation}
Known Clifford constructions of the electric-magnetic exchange in the $\mathbb{Z}_2$ toric code instead have order four on the full physical Hilbert space, even though their action on the anyon types has order two~\cite{Hastings:2021ptn, Barkeshli:2022wuz, Kobayashi2024crosscap, Tu:2025bqf, Shirley:2025yji}.
Moreover, the extra microscopic action can be removed by an appropriate dressing only at the cost of introducing non-Clifford operations, yielding an exact order-two realization~\cite{Tu:2025bqf}.
This suggests that the mismatch is not accidental, but reflects an obstruction intrinsic to the Clifford structure.
Indeed, a recent work by Kobayashi established such an obstruction under additional assumptions, including translation invariance with restricted families of system sizes~\cite{Kobayashi:2026upe}.

In this paper, we rigorously prove that Clifford realizations of the electric-magnetic exchange in the $\mathbb Z_2$ toric code cannot be order two. 
Our result does not assume translation symmetry or a fixed pairing between vertex and plaquette stabilizers.
More generally, an individual stabilizer may be mapped to an arbitrary local product of stabilizers of the opposite type.
Our proof uses a particular arrangement of electric and magnetic string operators, chosen so that one electric--magnetic pair crosses once near the middle and therefore has nontrivial mutual braiding.
Geometrically, this string configuration locally reproduces the unavoidable crossing associated with a cross-cap, while the system itself remains on an ordinary orientable lattice.

In the toric code, Clifford operations have a particularly natural physical interpretation.
Electric and magnetic excitations are created by open string operators, and the full Pauli group is generated by these strings and their products.
Thus, Clifford operations can be understood as transformations that map string operators to other string operators, possibly including their products.
From this perspective, the obstruction can be understood as a problem of extending electric-magnetic duality from closed strings to open strings.
Electric-magnetic duality pairs vertex and plaquette stabilizers, which can be regarded as elementary contractible closed-string operators.
The question is whether this order-two exchange can be extended consistently to the entire set of open-string operators.
An order-two Clifford realization would require a corresponding order-two transformation of all open-string operators.
Our result shows that such an extension is impossible.

We further study the same question for the $\mathbb{Z}_N$ toric code and find that the obstruction is governed entirely by the parity of $N$.
For odd $N$, the order-two electric-magnetic duality admits an order-two Clifford realization, whereas for even $N$ such a realization is impossible.
We also construct an order-four Clifford realization for every even $N$, showing that the minimum Clifford order is exactly four.
By contrast, once non-Clifford operations are allowed, an exact order-two electric-magnetic exchange can be realized for arbitrary $N$.

\section{$e$-$m$ exchange on the $\mathbb{Z}_2$ toric code} \label{sec:z2-toric-code}

In this section, we prove that no locality-preserving Clifford realization of the $e$-$m$ exchange can have order two.

\subsection{Microscopic $e$-$m$ exchange}

We begin by recalling the $\mathbb{Z}_2$ toric code and fixing our conventions.
Let $\Lambda$ be a square lattice on a closed orientable surface, with vertex, edge, and plaquette sets $V(\Lambda)$, $E(\Lambda)$, and $F(\Lambda)$, respectively.
A qubit is placed on each edge $e\in E(\Lambda)$, and the toric-code Hamiltonian is
\begin{align} \label{eq:toric-code-hamiltonian}
H_{\mathrm{TC}}:=\;-\sum_{v\in V(\Lambda)}A_v\ -\sum_{p\in F(\Lambda)}B_p,\qquad 
A_v:=
\figbox{0.9}{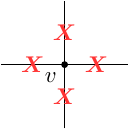}
,\quad B_p:=
\figbox{0.9}{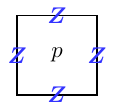}
.
\end{align}
A ground state is a simultaneous $+1$ eigenstate of all stabilizers. $A_v|\psi\rangle=B_p|\psi\rangle=|\psi\rangle$.
On a torus, these stabilizer conditions satisfy two global redundancy constraints,
\begin{align}
\prod_{v\in V(\Lambda)}A_v=1,\qquad
\prod_{p\in F(\Lambda)}B_p=1.
\end{align}
The ground space on the torus is four-dimensional.

\begin{figure}[!t]
\centering
\raisebox{-79.98862pt}[84.70319pt][79.98862pt]{%
\includegraphics[width=491.43622pt,height=164.69181pt]{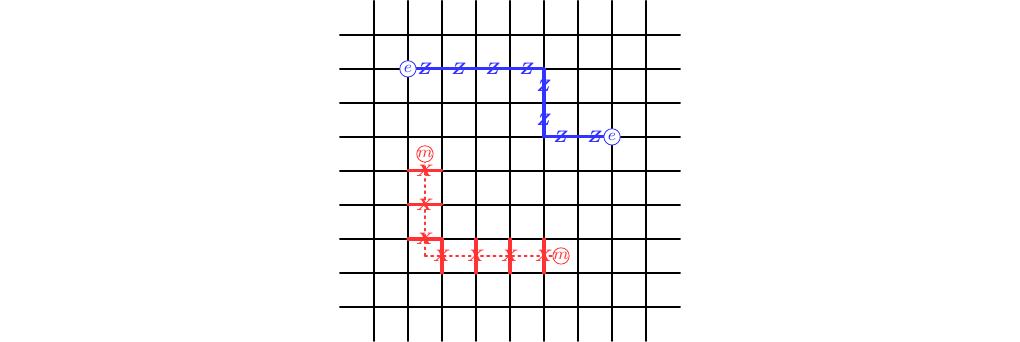}}
\caption{Anyon excitations in $\mathbb{Z}_2$ toric code}
\label{fig:toric-cod-anyon-strings}
\end{figure}

Anyonic excitations of the toric code can be characterized by open string operators.
Let $\gamma_Z$ and $\gamma_X$ denote open strings composed of Pauli $Z$ and $X$ operators, respectively.
For a ground state $|\psi\rangle$, the excited state $\gamma_Z|\psi\rangle$ violates the stabilizer conditions associated with the vertex operators $A_v$ at the two endpoints of $\gamma_Z$, thereby creating a pair of $e$ anyons.
Similarly, $\gamma_X|\psi\rangle$ violates the plaquette stabilizers $B_p$, producing a pair of $m$ anyons (see Fig.\,\ref{fig:toric-cod-anyon-strings}).


Previous works have shown that the $e$-$m$ exchange can be realized by an \emph{order-four} finite-depth local \emph{Clifford} circuit.
One such realization is the following depth-$3$ unitary~\cite{Hastings:2021ptn, Barkeshli:2022wuz, Tu:2025bqf, Shirley:2025yji},
\begin{align} \label{eq:microscopic-em-exchange}
U_{em}:=\prod_{p\in F(\Lambda)}e^{\frac{i\pi}{4}P_3(p)}\prod_{q\in F(\Lambda)}e^{-\frac{i\pi}{4}P_2(q)}\prod_{r\in F(\Lambda)}e^{\frac{i\pi}{4}P_1(r)},
\end{align}
where the two-qubit Pauli operators are
\begin{align}
P_1(p):=\;
\figbox{0.9}{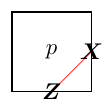}
,\quad P_2(p):=\;
\figbox{0.9}{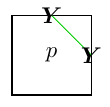}
,\quad P_3(p):=\;
\figbox{0.9}{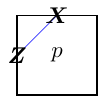}
.
\end{align}
We can indeed verify that the paired vertex and plaquette stabilizers are exchanged,
\begin{align}
U_{em}A_vU_{em}^{\dagger}=B_{p(v)},\qquad
U_{em}B_{p(v)}U_{em}^{\dagger}=A_v,
\label{eq:em-action-on-tric-code-stabilizers}
\end{align}
where $p(v)$ denotes the plaquette immediately to the upper right of $v$, as shown explicitly below.\footnote{
To determine how $U_{em}$ transforms the stabilizer generators and how it acts on the anyon labels, it is useful to recall a simple Clifford conjugation rule.
For any Pauli string $P$, the unitary
\begin{align}
U:=e^{\frac{i\pi}{4}P}
=\frac{1}{\sqrt{2}}(1+iP)
\end{align}
is a Clifford gate whose action on a Pauli string $Q$ is
\begin{align}
UQU^{\dagger}=
\begin{cases}
Q & ([P,Q]=0), \\
iPQ & (\{P,Q\}=0).
\end{cases}
\label{eq:clifford-conjugation-rule}
\end{align}
}
\begin{equation} \label{eq:em-action-star-plaquette-picture}
\begin{gathered}
\figbox{1}{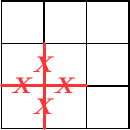}
\;\xrightarrow{\,e^{\frac{i\pi}{4}P_1(p)}\,}\;
\figbox{1}{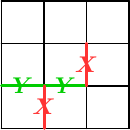}
\;\xrightarrow{\,e^{-\frac{i\pi}{4}P_2(p)}\,}\;
\figbox{1}{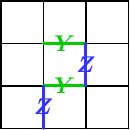}
\;\xrightarrow{\,e^{\frac{i\pi}{4}P_3(p)}\,}\;
\figbox{1}{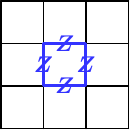}
\,,
\\[1em]
\figbox{1}{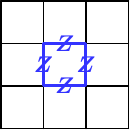}
\;\xrightarrow{\,e^{\frac{i\pi}{4}P_1(p)}\,}\;
\figbox{1}{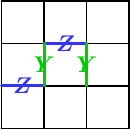}
\;\xrightarrow{\,e^{-\frac{i\pi}{4}P_2(p)}\,}\;
\figbox{1}{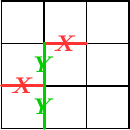}
\;\xrightarrow{\,e^{\frac{i\pi}{4}P_3(p)}\,}\;
\figbox{1}{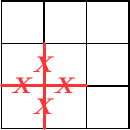}
\,.  
\end{gathered} 
\end{equation}

Since $U_{em}$ exchanges the paired vertex and plaquette stabilizers, one may be tempted to speculate that the microscopic unitary itself also has order two.
However, this is not the case.
A direct calculation shows that
\begin{align}
U_{em}^4\propto I,
\end{align}
while $U_{em}^2$ is nontrivial on the physical Hilbert space.
\begin{figure}
\centering
\raisebox{-54.45436pt}[309.88889pt][54.45436pt]{%
\includegraphics[width=491.43622pt,height=364.34325pt]{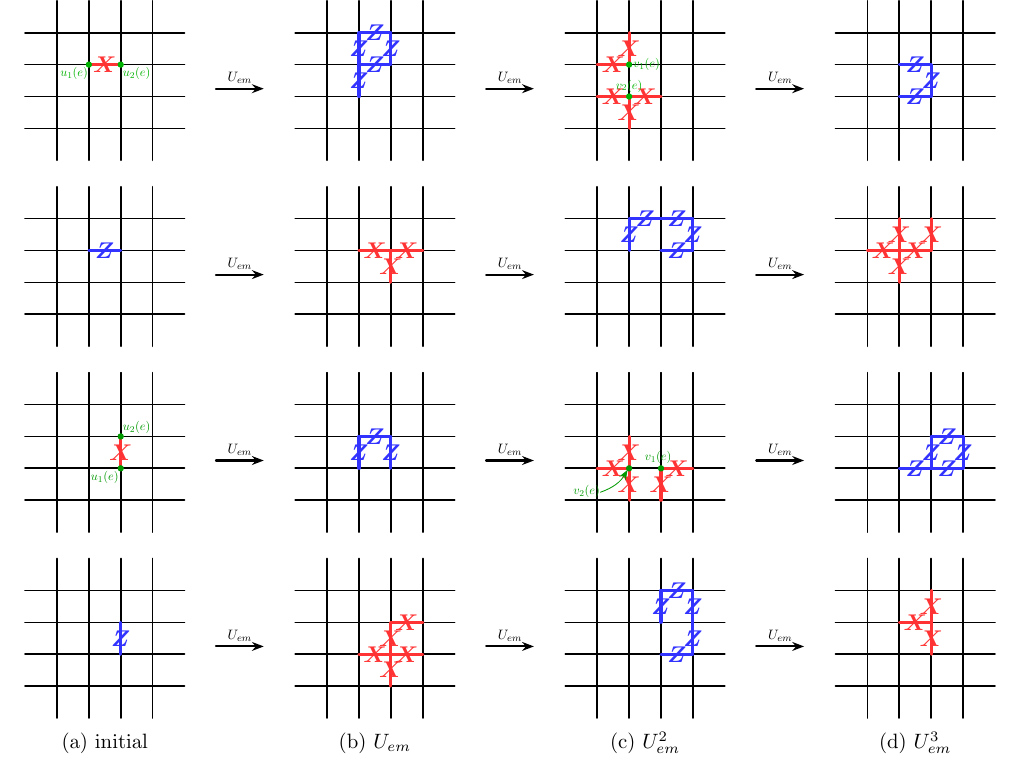}}
\caption{Conjugation rules for the explicit $\mathbb{Z}_2$ exchange $U_{em}$.
(a)--(d) display the four successive images of single-edge Pauli operators under repeated conjugation by $U_{em}$.
Red and blue labels denote $X$- and $Z$-type factors, respectively.}
\label{fig:u-action}
\end{figure}
The origin of this order-four behavior can be clearly seen from the conjugation of local Pauli operators shown in Fig.\,\ref{fig:u-action}.
For the single-edge Pauli operators $P$ shown in Fig.\,\ref{fig:u-action}, one can observe that two successive applications of $U_{em}$ do not return $P$ to itself.
Instead, one finds schematically
\begin{align}
U_{em}^2 P U_{em}^{-2} = P S_P ,
\end{align}
where $S_P$ is a nontrivial product of toric-code stabilizers.
In fact, this behavior can be established rigorously.
For any Pauli operator $P$ supported in a simply connected finite region, $U_{em}^2$ returns $P$ to itself, up to a phase, if and only if $P$ belongs to the toric-code stabilizer group.
We prove this statement in Appendix~\ref{app:2vs4}.

This naturally raises the question of whether the order-four behavior is a special feature of the particular circuit in Eq.\,\eqref{eq:microscopic-em-exchange}, or whether it is unavoidable for any Clifford realization of the $e$-$m$ exchange.
Our main result shows that the latter is the case.

\begin{theorem} \label{thm:clifford-em-not-z2}
Consider the $\mathbb{Z}_2$ toric code on an $L_x\times L_y$ square lattice with periodic boundary conditions, with the system size sufficiently large compared with the locality range.
Any locality-preserving Clifford realization $U_{em}$ of the $e$-$m$ exchange satisfies
\begin{align}
U_{em}^2 \not\propto I .
\end{align}
In particular, there is no order-two Clifford realization of the $e$-$m$ exchange.
\end{theorem}

\paragraph{Remark on previous work.}
A related no-go result was recently obtained in ref.\,\cite{Kobayashi:2026upe}, but only under substantial restrictions.
The rigorous result there applies to implementations with an enlarged $l\times l$ translation symmetry for odd $l$.
Moreover, the $e$-$m$ exchange is restricted to diagonal pairings only.
These restrictions are closely tied to the polynomial formalism for translation-invariant Pauli systems where the obstruction is expressed as an abstract algebraic incompatibility.
While this establishes the no-go result within that setting, it does not reveal the physical origin of the obstruction, e.g. the role of the braiding structure of $e$ and $m$ in the no-go mechanism.

Our result fully removes these restrictions.
We do not assume any translation symmetry, and the theorem applies to arbitrary $L_x\times L_y$ periodic square lattices.
We also allow a much more general microscopic realization of the $e$-$m$ exchange than a fixed pairwise mapping between vertex and plaquette stabilizers.
In particular, a vertex operator may be mapped to an arbitrary local product of plaquette stabilizers, and a plaquette operator may similarly be mapped to an arbitrary local product of vertex stabilizers, provided that these transformations are local and locally invertible.
Furthermore, our proof makes the role of $e$-$m$ braiding statistics in the obstruction physically transparent.

\paragraph{Order-two non-Clifford realization.}
An exact order-two realization of the $e$-$m$ exchange can be obtained by dressing the Clifford unitary $U_{em}$ in Eq.\,\eqref{eq:microscopic-em-exchange} with a non-Clifford operator \cite{Tu:2025bqf},
\begin{align}
\widetilde{U}_{em}:=U_{em}D,
\qquad
D:=\prod_v
i^{\frac{1-A_v}{2}\frac{1-B_{p(v)}}{2}},
\qquad
\widetilde{U}_{em}^2\propto I.
\label{eq:z2-non-clifford-dressing}
\end{align}
The dressing $D$ assigns a phase $i$ whenever both the electric and magnetic excitations are present at a paired vertex and plaquette, corresponding to the composite fermion $f=em$.
Intuitively, applying the $e$-$m$ exchange twice returns the anyons to their original types but produces a nontrivial braiding phase when both are present.
The non-Clifford dressing precisely cancels this phase and restores an exact order-two symmetry.

\subsection{Obstruction for diagonal exchange}

We first prove the obstruction under the simplifying assumption that $U$ exchanges each vertex operator $A_v$ with its diagonally paired plaquette operator $B_p$, as in Eq.\,\eqref{eq:em-action-on-tric-code-stabilizers}.
Throughout this subsection, $p(v)$ denotes the plaquette immediately to the upper right of the vertex $v$.
The general case, in which vertex and plaquette stabilizers may be mapped to local products of stabilizers, will be considered in the next subsection.

\begin{theorem} \label{thm:clifford-em-not-z2-diagonal}
Consider the $\mathbb{Z}_2$ toric code on a square lattice $\Lambda$ with periodic boundary conditions, with the system size sufficiently large compared with the locality range.
If a locality-preserving Clifford unitary $U$ induces diagonal exchanges of vertex and plaquette stabilizers
\begin{align*}
UA_vU^\dagger=B_{p(v)},
\qquad
UB_{p(v)}U^\dagger=A_v
\end{align*}
for every $v\in V(\Lambda)$, then $U^2\not\propto I$.
\end{theorem}

We will use the following useful property of string operators in the toric code.

\begin{lemma}[Local stabilizer decomposition]
\label{lem:diagonal-exchange-string-dressing}
Let $U$ satisfy the exchange relations in Theorem~\ref{thm:clifford-em-not-z2-diagonal}.
Let $\alpha_e$ be an open Pauli-$Z$ string connecting vertices $v_1,v_2$, and let $\alpha_m$ be a nearby Pauli-$X$ string connecting the paired plaquettes $p(v_1),p(v_2)$.
Choose a disk-shaped neighborhood $R_{\alpha_e}$ of the strings, wide enough to contain the support of $U\alpha_eU^\dagger$.
Then
\begin{align}
U\alpha_eU^\dagger\propto\alpha_m S_{R_{\alpha_e}},
\label{eq:diagonal-local-string-dressing}
\end{align}
where $S_{R_{\alpha_e}}$ has a unique decomposition into vertex and plaquette stabilizers in $R_{\alpha_e}$.
\end{lemma}

The analogous statement holds with electric and magnetic strings interchanged.

\begin{proof}
Since $U$ is Clifford, $U\alpha_eU^\dagger$ is a Pauli operator.
It creates the same pair of magnetic excitations as $\alpha_m$, so $U\alpha_eU^\dagger\alpha_m$ commutes with every stabilizer.
This operator is supported in a disk-topology region $R_{\alpha_e}$ and hence is proportional to a product of stabilizers in $R_{\alpha_e}$ whose decomposition is unique.
The global redundancy relations give alternative decompositions, but these involve stabilizer generators outside $R_{\alpha_e}$.
\end{proof}

\begin{proof}[Proof of Theorem \ref{thm:clifford-em-not-z2-diagonal}]
Suppose, for contradiction, that $U^2\propto I$.
After absorbing an overall phase into $U$, we may assume $U^2=I$.
We choose three vertex--plaquette pairs
\begin{align}
h_1=(v_1,p_1),\qquad h_2=(v_2,p_2),\qquad h_3=(v_3,p_3),
\qquad p_j=p(v_j),
\label{eq:diagonal-three-pairs}
\end{align}
in the L-shaped arrangement shown in Fig.\,\ref{fig:physical-picture-geometry}, with separations much larger than the locality range of $U$.
Let $\alpha_e$ and $\alpha_m$ be electric and magnetic strings connecting $h_1$ to $h_2$, and let $\beta_e$ and $\beta_m$ connect $h_2$ to $h_3$.
We choose their paths so that $\alpha_m$ and $\beta_e$ cross once, whereas $\alpha_e$ and $\beta_m$ are disjoint.

\begin{figure}[!t]
\centering
\makebox[\textwidth][c]{%
\includegraphics[scale=1.4]{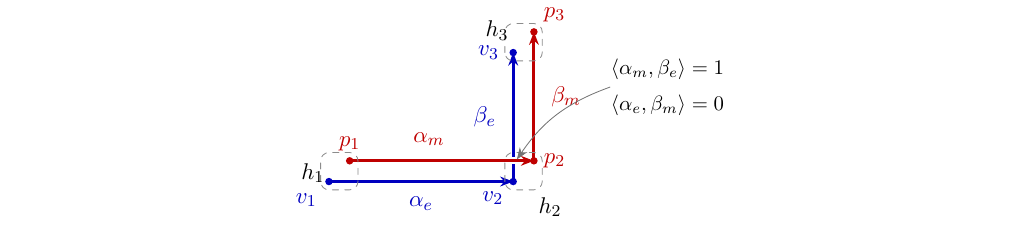}}
\caption{String geometry used in the proof of Theorem \ref{thm:clifford-em-not-z2-diagonal}.
Blue electric strings lie on the direct lattice and red magnetic strings lie on the dual lattice.}
\label{fig:physical-picture-geometry}
\end{figure}

By Lemma \ref{lem:diagonal-exchange-string-dressing}, $U\alpha_eU^\dagger$ is the magnetic string $\alpha_m$ decorated by a local product of stabilizers.
Here, we separate the factors $A_{v_2}$ and $B_{p_2}$, associated with $h_2=(v_2,p_2)$, from the remaining stabilizers.
\begin{align}
U\alpha_eU^\dagger
\propto\alpha_m A_{v_2}^{x}B_{p_2}^{y}\widetilde{S}_{\alpha_e},
\qquad x,y\in\mathbb{F}_2,
\label{eq:diagonal-alpha-image}
\end{align}
where $\widetilde{S}_{\alpha_e}$ contains neither $A_{v_2}$ nor $B_{p_2}$.
These factors $A_{v_2}$ and $B_{p_2}$ determine how the stabilizer decoration contributes to the commutation relations with the $\beta$ strings.

Applying $U$ once more gives
\begin{align}
\alpha_e
\propto
(U\alpha_mU^\dagger)B_{p_2}^{x}A_{v_2}^{y}
(U\widetilde{S}_{\alpha_e}U^\dagger).
\label{eq:diagonal-alpha-twice}
\end{align}
Since $U$ exchanges the stabilizers pairwise, $U\widetilde{S}_{\alpha_e}U^\dagger$ also contains neither $A_{v_2}$ nor $B_{p_2}$.
Hence, the vertex and plaquette stabilizers at $h_2=(v_2,p_2)$ have their exponents interchanged. $A_{v_2}^{x}B_{p_2}^{y} \rightarrow B_{p_2}^{x}A_{v_2}^{y}$.
The same reasoning applies to the $\beta$ strings, and we obtain
\begin{align}
U\alpha_eU^\dagger
&\propto\alpha_m A_{v_2}^{x}B_{p_2}^{y}\widetilde{S}_{\alpha_e},
\nonumber\\
U\alpha_mU^\dagger
&\propto\alpha_e A_{v_2}^{y}B_{p_2}^{x}\widetilde{S}_{\alpha_m},
\nonumber\\
U\beta_eU^\dagger
&\propto\beta_m A_{v_2}^{z}B_{p_2}^{w}\widetilde{S}_{\beta_e},
\nonumber\\
U\beta_mU^\dagger
&\propto\beta_e A_{v_2}^{w}B_{p_2}^{z}\widetilde{S}_{\beta_m},
\qquad x,y,z,w\in\mathbb{F}_2.
\label{eq:physical-picture-string-images}
\end{align}
Each tilded operator is a product of stabilizers near its corresponding string and contains neither $A_{v_2}$ nor $B_{p_2}$.
Overall Pauli phases will not affect the argument.

For Pauli operators $P,Q$, define the commutation relation by
\begin{align}
PQ=(-1)^{\langle P,Q\rangle}QP,
\qquad \langle P,Q\rangle\in\mathbb{F}_2.
\label{eq:diagonal-pauli-pairing}
\end{align}
Since $\alpha_m$ and $\beta_e$ cross once, we have
\begin{align}
1
&=\langle\alpha_m,\beta_e\rangle
=\langle U\alpha_mU^\dagger,U\beta_eU^\dagger\rangle
\nonumber\\
&=\left\langle
\alpha_e A_{v_2}^{y}B_{p_2}^{x},
\beta_m A_{v_2}^{z}B_{p_2}^{w}
\right\rangle
\nonumber\\
&=z\langle\alpha_e,A_{v_2}\rangle
+x\langle B_{p_2},\beta_m\rangle
=x+z.
\label{eq:diagonal-commutation-calculation}
\end{align}
In the second line we dropped the tilded stabilizers as they do not contain $A_{v_2}$ or $B_{p_2}$.
In the third line we used that $\alpha_e$ and $\beta_m$ are disjoint, while $\alpha_e$ anticommutes with $A_{v_2}$ and $\beta_m$ anticommutes with $B_{p_2}$.
Thus
\begin{align}
x+z=1\pmod{2}.
\label{eq:physical-picture-crossing}
\end{align}

We now derive another constraint by closing the strings into contractible loops.
The product $\alpha_e\beta_e$ is an electric string connecting $v_1$ and $v_3$ (Fig.\,\ref{fig:no-go-string-geometry}(a)).
Multiplying the transformations in Eq.\,\eqref{eq:physical-picture-string-images}, we find
\begin{align}
U(\alpha_e\beta_e)U^\dagger
&\propto\alpha_m\beta_m A_{v_2}^{a}B_{p_2}^{b}
\widetilde{S}_{\alpha_e\beta_e},
\nonumber\\
a&=x+z,\qquad b=y+w\pmod{2},
\label{eq:diagonal-joined-string-image}
\end{align}
where $\widetilde{S}_{\alpha_e\beta_e}=\widetilde{S}_{\alpha_e}\widetilde{S}_{\beta_e}$ contains neither middle generator.

Choose an electric string $\gamma_e$ from $v_3$ back to $v_1$, following the distant sides of the rectangle in Fig.\,\ref{fig:no-go-string-geometry}(b).
Here, $\alpha_e\beta_e\gamma_e$ forms a closed loop surrounding a rectangular region $M$, and can be expressed as a product of plaquettes on $M$
\begin{align}
\alpha_e\beta_e\gamma_e=\prod_{q\in M}B_q.
\label{eq:diagonal-electric-loop}
\end{align}
The region $M$ lies to the left of $\beta_e$, whereas $p_2$ lies immediately to the upper right of $v_2$. Hence, 
\begin{align}
p_2\notin M.
\end{align}
Using the exchange relations of stabilizers, we have
\begin{align}
U(\alpha_e\beta_e\gamma_e)U^\dagger
=\prod_{v\in V_M}A_v,
\qquad
V_M:=\{v\in V(\Lambda):p(v)\in M\}.
\label{eq:diagonal-electric-loop-image}
\end{align}
Since $p(v)$ is the plaquette immediately to the upper right of $v$, the vertex region $V_M$ is obtained from the plaquette region $M$ by shifting it to the lower left.
This direction is opposite to the vertex-to-plaquette pairing $A_v\mapsto B_{p(v)}$, which shifts the stabilizer label to the upper right.
In particular, since $p_2\notin M$, we have
\begin{align}
v_2\notin V_M.
\end{align}

\begin{figure}[t]
\centering
\raisebox{-18.0pt}[234.59846pt][18.0pt]{%
\includegraphics[width=491.43622pt,height=252.59846pt]{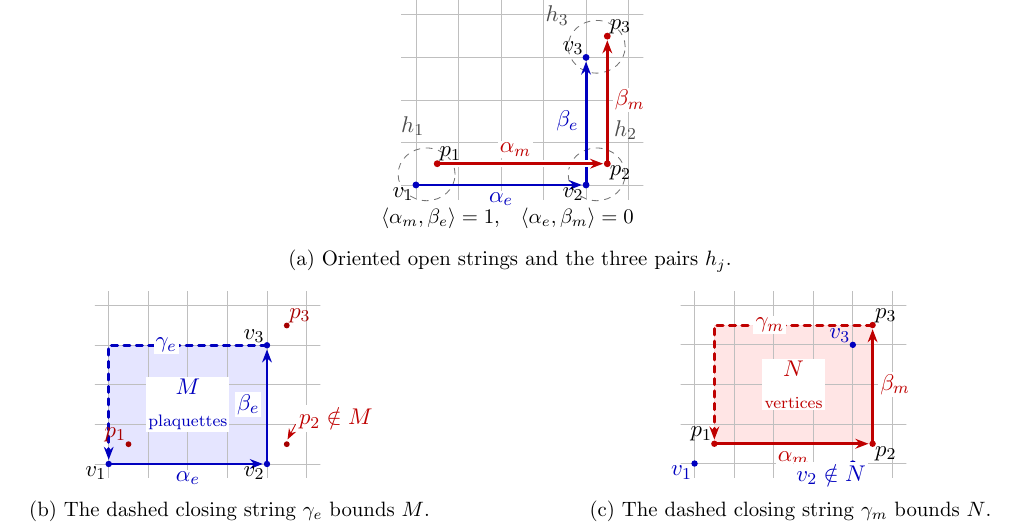}}
\caption{String geometry used in the proof of Theorem \ref{thm:clifford-em-not-z2-diagonal}.
Blue direct-lattice strings are Pauli-$Z$ operators and red dual-lattice strings are Pauli-$X$ operators; arrows give the orientations used for the $\mathbb{Z}_N$ argument.
(a) $\alpha_m$ and $\beta_e$ cross once, while $\alpha_e$ and $\beta_m$ are disjoint.
Dashed closing strings form (b) an electric loop enclosing the plaquette set $M$ with $p_2\notin M$, and (c) a magnetic loop enclosing the vertex set $N$ with $v_2\notin N$.}
\label{fig:no-go-string-geometry}
\end{figure}

Likewise, choose a nearby magnetic closing string $\gamma_m$ from $p_3$ to $p_1$, as in Fig.\,\ref{fig:no-go-string-geometry}(c).
The loop $\alpha_m\beta_m\gamma_m$ bounds a simply connected vertex region $N$, and thus
\begin{align}
\alpha_m\beta_m\gamma_m=\prod_{v\in N}A_v.
\label{eq:diagonal-magnetic-loop}
\end{align}
Here, $N$ and $V_M$ do not coincide.
Since $p(v)$ lies immediately to the upper right of $v$, $V_M$ is shifted to the lower left relative to $N$.
In particular, the lower boundary of $N$ is $\alpha_m$, which lies above $v_2$, and hence
\begin{align}
v_2\notin N.
\end{align}

The closing string $\gamma_e$ can be chosen sufficiently far from the middle pair $(v_2,p_2)$.
Hence, Lemma \ref{lem:diagonal-exchange-string-dressing} gives
\begin{align}
U\gamma_eU^\dagger\propto\gamma_m\widetilde{S}_{\gamma_e},
\label{eq:diagonal-closing-string-image}
\end{align}
where $\widetilde{S}_{\gamma_e}$ contains neither $A_{v_2}$ nor $B_{p_2}$.

We can now compute the transformed electric loop in a second way, by multiplying the three transformed segments.
Combining Eqs.\,\eqref{eq:diagonal-joined-string-image}, \eqref{eq:diagonal-electric-loop-image}, and \eqref{eq:diagonal-closing-string-image}, we obtain
\begin{align}
\alpha_m\beta_m\gamma_m
A_{v_2}^{a}B_{p_2}^{b}
\widetilde{S}_{\alpha_e\beta_e}\widetilde{S}_{\gamma_e}
\propto\prod_{v\in V_M}A_v.
\label{eq:diagonal-loop-comparison}
\end{align}
Substituting Eq.\,\eqref{eq:diagonal-magnetic-loop} gives an equality between stabilizer products,
\begin{align}
\left(\prod_{v\in N}A_v\right)
A_{v_2}^{a}B_{p_2}^{b}
\widetilde{S}_{\alpha_e\beta_e}\widetilde{S}_{\gamma_e}
\propto\prod_{v\in V_M}A_v.
\label{eq:diagonal-stabilizer-comparison}
\end{align}
Neither $N$ nor $V_M$ contains $v_2$, and the tilded products contain no $A_{v_2}$.
The only possible occurrence of this generator is therefore the explicit factor $A_{v_2}^{a}$ on the left-hand side.
Comparing its coefficient in the two disk-supported decompositions yields
\begin{align}
a=x+z=0\pmod{2},
\label{eq:physical-picture-loop}
\end{align}
in contradiction with Eq.\,\eqref{eq:physical-picture-crossing}.
This completes the proof.
\end{proof}

The proof suggests a possible connection with the real projective plane $\mathbb{RP}^2$ and cross-caps \cite{Kobayashi2024crosscap}.
Recall that $\mathbb{RP}^2$ can be obtained from a disk by identifying antipodal points on its boundary.
Namely, a path connecting two antipodal boundary points becomes a closed loop.
A striking feature of this geometry is that two such paths with different endpoints must cross an odd number of times, even though the two loops can be continuously deformed into one another.
Thus, the geometry forces a crossing that cannot be removed by deforming the loops.
In our proof, the electric and magnetic strings are chosen to cross once near the middle pair.
This is analogous to the unavoidable crossing on $\mathbb{RP}^2$.
Our construction does not actually change the topology of the lattice.
Rather, the string configuration locally emulates a cross-cap-like
geometry and probes the response of the string operators to this geometry.

\subsection{Obstruction for general $e$-$m$ exchange}

The argument in the previous subsection assumed a particularly simple $e$-$m$ exchange pattern, in which each vertex operator $A_v$ is exchanged with the plaquette operator $B_p$ immediately to its upper right.
In this subsection, we extend the result to a much more general class of $e$-$m$ exchanges.

\begin{definition}[$e$-$m$ exchanging symmetry]
\label{def:z2-em-exchanging-symmetry}
A locality-preserving Clifford unitary $U$ is called an $e$-$m$ exchanging symmetry if there exists a constant $R$, independent of the system size, such that for every vertex $v$ and plaquette $p$,
\begin{align}
\begin{aligned}
U A_v U^\dagger
&=
\prod_{q\in \mathcal{P}(v)} B_q,
\\
U B_p U^\dagger
&=
\prod_{w\in \mathcal{V}(p)} A_w,
\end{aligned}
\label{eq:general-em-exchange}
\end{align}
where every plaquette in $\mathcal{P}(v)$ lies within distance $R$ from $v$, and every vertex in $\mathcal{V}(p)$ lies within distance $R$ from $p$.
\end{definition}

Thus, an individual vertex stabilizer may be mapped to an arbitrary local product of plaquette stabilizers.
No translation symmetry or fixed pairing between vertices and plaquettes is assumed.
Not every choice of the sets $\mathcal{P}(v)$ and $\mathcal{V}(p)$ defines a physically consistent transformation, as it must define an invertible transformation between the physical vertex and plaquette stabilizer groups.

Our goal is to show that the no-go theorem continues to hold for this generalized notion of $e$-$m$ exchange.
The strategy is to reduce the general transformation Eq.\,\eqref{eq:general-em-exchange} to the simple pairwise exchange studied in the previous subsection.
Here we will use the standard binary formalism for the $\mathbb{Z}_2$ toric code.
The relevant binary vector spaces are the vertex, edge, and plaquette spaces
\begin{align}
C_0
&:=
\bigoplus_{v\in V(\Lambda)}\mathbb{F}_2 v,
\qquad
C_1
:=
\bigoplus_{e\in E(\Lambda)}\mathbb{F}_2 e,
\qquad
C_2
:=
\bigoplus_{p\in F(\Lambda)}\mathbb{F}_2 p.
\end{align}
For an edge vector $c \in C_1$, we denote the corresponding Pauli operators by
\begin{align}
X(c):=\prod_e X_e^{c_e},
\qquad
Z(c):=\prod_e Z_e^{c_e}.
\end{align}
The CSS chain complex of the toric code is
\begin{align}
C_0
\overset{G_1}{\longrightarrow}
C_1
\overset{G_2^t}{\longrightarrow}
C_2
\end{align}
where $G_1:C_0\longrightarrow C_1$ and $G_2:C_2\longrightarrow C_1$ are the vertex and plaquette maps.
In particular, $G_1v$ is the binary edge vector supported on the edges incident on $v$, while $G_2p$ is the binary edge vector supported on the boundary of $p$.
The toric-code stabilizers can be written as
\begin{align}
A_v=X(G_1v),
\qquad
B_p=Z(G_2p).
\end{align}
The commutativity of the vertex and plaquette stabilizers is expressed as
\begin{align}
G_2^tG_1=0
\end{align}
where $t$ denotes the transpose.

One can also extend this notation to arbitrary stabilizer products.
For $a\in C_0$ and $b\in C_2$, define
\begin{align}
A(a):=X(G_1a),
\qquad
B(b):=Z(G_2b).
\end{align}
Defining the all-ones plaquette vector
\begin{align}
\mathbf{1}_2
:=
\sum_{p\in F(\Lambda)}p\in C_2,
\end{align}
the global plaquette redundancy $\prod_p B_p=I$ can be written as
\begin{align}
B(\mathbf{1}_2)
=
Z(G_2\mathbf{1}_2)
=
I,
\qquad
G_2\mathbf{1}_2=0.
\end{align}

We now consider a local change of basis among the plaquette stabilizers.
Such a transformation can be represented by an invertible $\mathbb{F}_2$-linear map $F$.
\begin{align}
B(b)\longmapsto B(Fb), \qquad F:C_2\longrightarrow C_2.
\end{align}
Since an invertible change of stabilizer basis preserves the global redundancy $B(\mathbf{1}_2)=I$, we must have
\begin{align}
F\mathbf{1}_2=\mathbf{1}_2.
\end{align}

The following lemma shows that every such local change of plaquette stabilizer basis can be realized microscopically by a locality-preserving CSS Clifford transformation.

\begin{lemma}[Local stabilizer basis change]
\label{lem:local-stabilizer-basis-change}

Let $F:C_2\longrightarrow C_2$ be a locality-preserving and locally invertible $\mathbb{F}_2$-linear map satisfying $F\mathbf{1}_2=\mathbf{1}_2$.
Then there exists a locality-preserving CSS Clifford unitary $V_F$ such that
\begin{align}
\begin{aligned}
V_F A_v V_F^\dagger &= A_v,
\\
V_F B(b)V_F^\dagger &= B(Fb)
\end{aligned}
\label{eq:local-stabilizer-basis-change}
\end{align}
for every vertex $v$ and every $b\in C_2$.
\end{lemma}

Although we state the lemma for the square-lattice toric code, the proof does not rely on the square geometry.
The same idea is also expected to extend to broader classes of CSS stabilizer codes.

\begin{proof}
We first show that the deviation of $F$ from the identity can be written as
\begin{align}
L:= F-I=HG_2
\end{align}
for some locality-preserving linear map $H:C_1\to C_2$.

Since $F\mathbf{1}_2=\mathbf{1}_2$, we have
\begin{align}
L\mathbf{1}_2=(F-I)\mathbf{1}_2=0.
\label{eq:L-even-parity}
\end{align}
For each plaquette $p$, let $\ell_p$ denote the transpose of the $p$-th row of $L$.
Then, Eq.\,\eqref{eq:L-even-parity} implies
\begin{align}
\ell_p^t\mathbf{1}_2=0,
\end{align}
so $\operatorname{supp}(\ell_p)$ contains an even number of plaquettes.
Then, we can pair plaquettes in $\operatorname{supp}(\ell_p)$ as $(q_i,q_i')$.
For each pair, one may choose some simple sequence of adjacent plaquettes
\begin{align}
q_i=p_0,\ p_1,\ \ldots,\ p_k=q_i'
\end{align}
within the neighborhood of $p$.
Let $s_i$ be the edge vector supported on the edges shared by plaquettes in this sequence, as shown in Fig.\,\ref{fig:local-basis-change-path}.
\begin{figure}[t]
\centering
\raisebox{-0.0pt}[54.42168pt][0.0pt]{%
\includegraphics[width=491.43622pt,height=54.42168pt]{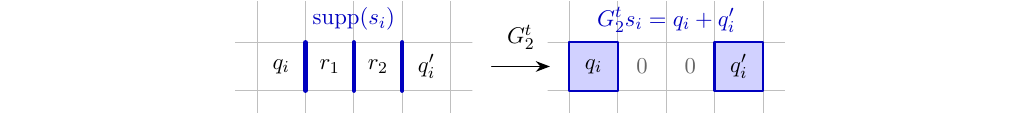}}
\caption{The edge vector $s_i$ (left) and the resulting plaquette vector $G_2^t s_i$ (right).}
\label{fig:local-basis-change-path}
\end{figure}
Each intermediate plaquette $p_j$ contains two selected edges from the sequence, while each endpoint plaquette contains only one.
Since we work over $\mathbb{F}_2$, the intermediate contributions cancel, and hence
\begin{align}
G_2^t s_i=q_i+q_i'.
\end{align}
Setting $h_p:=\sum_i s_i\in C_1$, we obtain
\begin{align}
G_2^t h_p
=\sum_i(q_i+q_i')
=\ell_p.
\end{align}
Taking $h_p^t$ as the $p$-th row of a locality-preserving map $H:C_1\to C_2$, we therefore have
\begin{align}
L=HG_2.
\label{eq:L-HG2}
\end{align}

We now lift the plaquette-label transformation $F$ to a locality-preserving invertible transformation on the physical edge labels.
Define
\begin{align}
T:=I+G_2H,
\qquad
T:C_1\longrightarrow C_1.
\end{align}
Then, $T$ reproduces the action of $F$ on plaquette stabilizers.
\begin{align}
TG_2
&=(I+G_2H)G_2 \nonumber\\
&=G_2+G_2HG_2 \nonumber\\
&=G_2+G_2(F-I)
=G_2F.
\end{align}
At the same time, $T$ leaves the vertex stabilizers invariant.
Namely, using the commutativity condition $G_2^tG_1=0$, we find
\begin{align}
T^tG_1
=(I+H^tG_2^t)G_1
=G_1.
\end{align}
Here, $T$ is invertible.
Indeed, the inverse is
\begin{align}
T^{-1}=I+G_2F^{-1}H.
\end{align}
To verify this, we use $HG_2=F-I$ and compute
\begin{align}
&(I+G_2H)(I+G_2F^{-1}H) \nonumber\\
&\qquad
=I+G_2H+G_2F^{-1}H+G_2HG_2F^{-1}H \nonumber\\
&\qquad
=I+G_2\!\left[
H+F^{-1}H+(F-I)F^{-1}H
\right]
=I.
\end{align}
Since $G_2$, $H$, and $F^{-1}$ are all locality preserving, both $T$ and $T^{-1}$ are locality preserving.

We now extend $T$ to a transformation of the full Pauli group.
Up to an overall phase, an edge Pauli operator can be labeled by a pair $(z,x)^t\in C_1\oplus C_1$, corresponding to
\begin{align}
Z(z)X(x).
\end{align}
Since the $Z$-type labels transform as $z\mapsto Tz$, preservation of the Pauli commutation relations requires the $X$-type labels to transform as $x\mapsto (T^{-1})^t x$.
We therefore define
\begin{align}
S:=
\begin{pmatrix}
T&0\\
0&(T^{-1})^t
\end{pmatrix}.
\end{align}
This transformation is symplectic and hence can be implemented by some Clifford unitary, which we denote by $V_F$.
Since $T$ and $T^{-1}$ are locality preserving,  both $S$ and $S^{-1}$ are locality preserving, and thus $V_F$ defines a locality-preserving CSS Clifford transformation.
Finally, one can verify
\begin{align}
V_F B_p V_F^\dagger
&= Z(TG_2p) \nonumber\\
&= Z(G_2Fp)
= B(Fp)
\end{align}
and
\begin{align}
V_F A_v V_F^\dagger
&= X\bigl((T^{-1})^tG_1v\bigr)\\
&= X\bigl(G_1v\bigr) = A_v.
\end{align}
\end{proof}

The same statement holds with vertices and plaquettes interchanged.

We are now ready to establish the general no-go result.
Suppose, for contradiction, that there exists an order-two Clifford $U$ which exchanges $e$ and $m$ with $U^2\propto I$ and
\begin{align}
U A(a) U^\dagger &= B(Fa), &
U B(b) U^\dagger &= A(Gb),
\end{align}
where $F$ and $G$ are locality-preserving invertible changes of stabilizer basis.
The condition $U^2\propto I$ implies that each column of $GF-I$ and $FG-I$ represents a trivial stabilizer product.
Since these columns are local, they cannot be nonzero all-ones vectors associated with the global redundancies, so $GF=FG=I$.
Thus $F^{-1}=G$ and $G^{-1}=F$ are also locality preserving.
Let $V_F$ be the CSS Clifford unitary as in Lemma~\ref{lem:local-stabilizer-basis-change}, and define
\begin{align}
\widehat U := V_F^\dagger U V_F .
\end{align}
Since $V_F$ fixes the vertex stabilizers and sends $B(b)$ to $B(Fb)$, we obtain
\begin{align}
\widehat U A(a)\widehat U^\dagger &= B(a), \\
\widehat U B(b)\widehat U^\dagger &= A(Kb),
\qquad K:=GF .
\end{align}
Therefore two applications of $\widehat U$ transform stabilizer generators as follows.
\begin{align}
\widehat U^2 A(a)\widehat U^{-2} &= A(Ka),\\
\widehat U^2 B(b)\widehat U^{-2} &= B(Kb).
\end{align}
Since $\widehat U$ is related to $U$ by conjugation, it also satisfies $\widehat U^2\propto I$.
Each column of $K-I$ therefore represents a trivial stabilizer product.
Since $K-I$ is local, no such column can be the nonzero all-ones vector associated with the global redundancy.
Hence,
\begin{align}
K=I.
\end{align}
Thus
\begin{align}
\widehat U A(a)\widehat U^\dagger=B(a),
\qquad
\widehat U B(b)\widehat U^\dagger=A(b),
\end{align}
reducing the general problem to the exact diagonal exchange studied in the previous subsection.
This completes the proof of Theorem~\ref{thm:clifford-em-not-z2}.

\section{$e$-$m$ exchange on the $\mathbb{Z}_N$ toric code} \label{sec:zn-toric-code}

We next generalize the no-go theorem to the $\mathbb{Z}_N$ toric code.
For odd $N$, the order-two electric--magnetic duality admits an order-two Clifford realization, whereas for even $N$ such a realization is impossible.

\begin{figure}[!t]
\centering
\raisebox{-87.67056pt}[92.38512pt][87.67056pt]{%
\includegraphics[width=491.43622pt,height=180.05568pt]{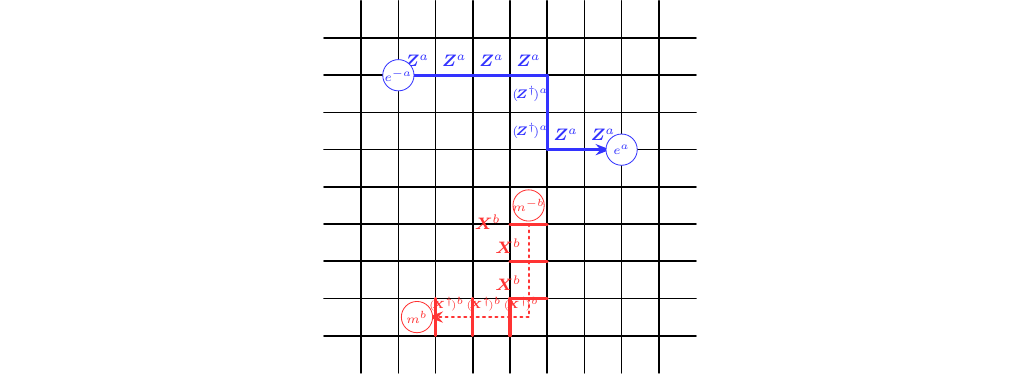}}
\caption{Anyon excitations in $\mathbb{Z}_N$ toric code.
The arrows indicate the orientation of the open $Z^a$-string and dual $X^b$-string.}
\label{fig:zn-toric-cod-anyon-strings}
\end{figure}

Consider $N$-dimensional qudits $\{|j\rangle\}_{j=0}^{N-1}$ and generalized Pauli operators $X$, $Z$ such that
\begin{align}
X|j\rangle=|j+1\rangle,\quad Z|j\rangle=\omega^j|j\rangle,\quad \omega:=e^{\frac{2\pi i}{N}}
\end{align}
where the basis label is understood modulo $N$.
Their commutation relation is $ZX=\omega XZ$.
The $\mathbb{Z}_N$ toric code Hamiltonian is given by
\begin{align} \label{eq:zn-toric-code-hamiltonian}
H:=\;-\sum_{v\in V(\Lambda)}A_v\ -\sum_{p\in F(\Lambda)}B_p,
\end{align}
where the vertex and plaquette projectors are
\begin{align} \label{eq:zn-toric-code-stabilizers}
A_v=\frac{1}{N}\sum_{k=0}^{N-1}A_v^{(k)},\quad B_p=\frac{1}{N}\sum_{k=0}^{N-1}B_p^{(k)},\quad A_v^{(k)}:=
\figbox{1}{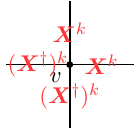}
,\quad B_p^{(k)}:=
\figbox{1}{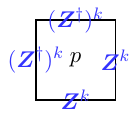}
.
\end{align}
An open oriented $Z^a$-string violates the vertex constraints at its two endpoints and creates a pair $e^a,e^{-a}$.
Similarly, an open dual $X^b$-string creates plaquette excitations $m^b,m^{-b}$ at its endpoints, as shown in Fig.\,\ref{fig:zn-toric-cod-anyon-strings}.

For an $e$-$m$ exchange, we consider the anyon permutation
\begin{align}
e\mapsto m^{-1},
\qquad
m\mapsto e^{-1},
\label{eq:ZN-em-exchange}
\end{align}
which exchanges the electric and magnetic sectors while preserving the anyon data.
For $N=2$, the inverses are invisible and this reduces to the usual exchange $e\leftrightarrow m$.

\subsection{Microscopic realization of the $e$-$m$ exchange anyon permutation}
\label{subsec:zn-setup}

We now construct an explicit Clifford realization of the $e$-$m$ exchange anyon permutation.
For Pauli operators $P$ and $Q$, define the commutation exponent
$\langle P,Q\rangle\in\mathbb{Z}_N$ by
\begin{align}
PQP^{\dagger}Q^{\dagger} = \omega^{\langle P,Q\rangle}.
\end{align}
Thus, $\langle P,Q\rangle=0$ if and only if $P$ and $Q$ commute.
In particular,
\[
\langle Z,X\rangle=-\langle X,Z\rangle=1.
\]

We first introduce a Clifford unitary associated with a Pauli operator $P$ satisfying $P^N=I$,
\begin{align}
U_{\bowtie}(P):=
\begin{cases}
\displaystyle
\frac{1}{\sqrt{N}}\sum_{r=0}^{N-1}\omega^{\frac{r(r+1)}{2}}P^r & (N:\text{odd}), \\[1.5em]
\displaystyle
\frac{1}{\sqrt{N}}\sum_{r=0}^{N-1}\omega^{\frac{r^2}{2}}P^r & (N:\text{even}).
\end{cases}
\end{align}
Its action on a Pauli operator $Q$ is, up to a $U(1)$ phase,
\begin{align}
Q\mapsto QP^{-\langle P,Q\rangle},
\end{align}
generalizing the qubit conjugation rule \eqref{eq:clifford-conjugation-rule}.

To verify this explicitly, let $m:=\langle P,Q\rangle$.
For odd $N$, direct computation gives
\begin{align}
\begin{aligned}
U_{\bowtie}(P)QU_{\bowtie}^\dagger(P)
&=\frac{1}{N}\sum_{r,s=0}^{N-1}
\omega^{\frac{1}{2}(r(r+1)-s(s+1))}\,P^rQ(P^\dagger)^s,\\
&=Q\cdot\frac{1}{N}\sum_{r,s=0}^{N-1}
\omega^{\frac{1}{2}(r(r+1)-s(s+1))+mr}\,P^{r-s},\\
&=Q\cdot\frac{1}{N}\sum_{t,s=0}^{N-1}
\omega^{\frac{t(t+1)}{2}+mt+(t+m)s}\,P^t,\\
&=\omega^{-\frac{m(m+1)}{2}}QP^{-m},
\end{aligned}
\label{eq:u_action}
\end{align}
where we used
\begin{align}
\sum_{s=0}^{N-1}\omega^{(t+m)s}=N\,\delta_{t,-m}.
\end{align}
Similarly, for even $N$,
\begin{align}
U_{\bowtie}(P)QU_{\bowtie}^\dagger(P)
=\omega^{-\frac{m^2}{2}}QP^{-m}.
\end{align}
Taking $Q=I$ also verifies that $U_{\bowtie}(P)$ is unitary.

We now use these elementary Clifford unitaries to construct a microscopic realization of the $e$-$m$ exchange.
For each plaquette $p$, define the weight-2 Pauli operators
\begin{align} \label{eq:zn-em-exchange-input-pauli}
P_1(p):=\;
\figbox{1}{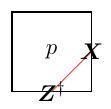}
,\quad P_2(p):=\;
\figbox{1}{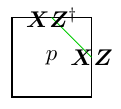}
,\quad P_3(p):=\;
\figbox{1}{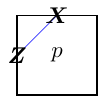}
.
\end{align}
Then define
\begin{align} \label{eq:zn-toric-code-em-exchange}
U_{em} := \prod_{p \in F(\Lambda)}U_{\bowtie}(P_3(p))
\prod_{q \in F(\Lambda)}U_{\bowtie}(P_2(q))
\prod_{r \in F(\Lambda)}U_{\bowtie}(P_1(r)).
\end{align}
Using the conjugation rule above, one finds that $U_{em}$ exchanges the $\mathbb{Z}_N$ toric-code stabilizer generators as
\begin{align}
\begin{aligned}
U_{em}A_v^{(1)}U_{em}^{\dagger}=B_{p(v)}^{(-1)},\quad
U_{em}B_{p(v)}^{(1)}U_{em}^{\dagger}=A_v^{(-1)}, \\
U_{em}A_v^{(-1)}U_{em}^{\dagger}=B_{p(v)}^{(1)},\quad
U_{em}B_{p(v)}^{(-1)}U_{em}^{\dagger}=A_v^{(1)}.
\end{aligned}
\label{eq:anyon_permutation_sym}
\end{align}
The corresponding transformations are illustrated in Fig.\,\ref{fig:u_action_qd}.
Thus, $U_{em}$ realizes the anyon permutation
\begin{align}
e\mapsto m^{-1},\qquad m\mapsto e^{-1}.
\end{align}
For $N=2$, this reduces to the microscopic $e$-$m$ exchange introduced in Section~2.

Although the induced anyon permutation has order two, the microscopic Clifford unitary $U_{em}$ generally has a higher order.
We determine its order below.

\begin{proposition} \label{prop:order-2n}
The order of $U_{em}$ is $2N$ up to an overall phase.
\end{proposition}

\begin{proof}
It follows from Eq.\,\eqref{eq:anyon_permutation_sym} that
\begin{align}
U_{em}^2A_v^{(1)}U_{em}^{-2}=A_v^{(1)},\qquad
U_{em}^2B_p^{(1)}U_{em}^{-2}=B_p^{(1)}.
\end{align}
Now take the horizontal edge $h=h_{0,0}$.
The conjugation action of $U_{em}^2$ on $X_h$ is (see Fig.\,\ref{fig:u_action_qd})
\begin{align}
U_{em}^2X_hU_{em}^{-2}
=
X_h\,A_{(0,0)}^{(1)}
\bigl(A_{(0,-1)}^{(1)}\bigr)^{-1}.
\end{align}
Since $U_{em}^2$ fixes every $A_v^{(1)}$, repeated application gives
\begin{align}
U_{em}^{2k}X_hU_{em}^{-2k}
=
X_h
\Bigl(
A_{(0,0)}^{(1)}
\bigl(A_{(0,-1)}^{(1)}\bigr)^{-1}
\Bigr)^k.
\end{align}
The stabilizer factor
$A_{(0,0)}^{(1)}\bigl(A_{(0,-1)}^{(1)}\bigr)^{-1}$
has order $N$.
Hence the conjugation action of $U_{em}^2$ has order at least $N$.
The same calculation for $Z_h$, $X_v$, and $Z_v$ shows that $U_{em}^{2N}$ acts trivially on the full Pauli algebra, and hence
\begin{align}
U_{em}^{2N}\propto I.
\end{align}
Moreover, the action on $X_h$ remains nontrivial for $0<k<N$, so no smaller even power can be proportional to the identity, while odd powers exchange vertex and plaquette stabilizers.
Therefore $U_{em}$ has order $2N$ up to an overall phase.

\end{proof}

\subsection{Obstruction to $\mathbb{Z}_2$ realization}

We now ask whether an order-two Clifford realization exists.
The answer is determined entirely by the parity of $N$.

When $N$ is odd, define
\begin{align}
V:=U_{em}^N.
\end{align}
Since $N$ is odd, $V$ induces the same $e$-$m$ exchange as $U_{em}$,
\begin{align}
VA_v^{(1)}V^{\dagger}=B_{p(v)}^{(-1)},\quad
VB_{p(v)}^{(1)}V^{\dagger}=A_v^{(-1)},\quad
VA_v^{(-1)}V^{\dagger}=B_{p(v)}^{(1)},\quad
VB_{p(v)}^{(-1)}V^{\dagger}=A_v^{(1)}.
\end{align}
By Proposition \ref{prop:order-2n}, $V$ has order two up to an overall phase.
Thus, there is no obstruction to an order-two Clifford realization for odd $N$.

For even $N$, write $N=2^aM$ with $M$ odd.
The unitary $U_{em}^M$ still realizes the $e$-$m$ exchange and has order $2^{a+1}$, but this does not rule out the existence of a different Clifford realization of lower order.

We now show that order two is impossible.

\begin{theorem} \label{thm:even-N-em-order-restated}
Consider the $\mathbb{Z}_N$ toric code on an $L_x\times L_y$ square lattice with periodic boundary conditions, with the system size sufficiently large compared with the locality range.
When $N$ is even, any locality-preserving Clifford realization $U_{em}$ of the $e$-$m$ exchange satisfies
\begin{align}
U_{em}^2 \not\propto I .
\end{align}
In particular, there is no order-two Clifford realization of the $e$-$m$ exchange.
\end{theorem}

The theorem establishes an obstruction to an order-two microscopic Clifford realization for even $N$.
Before turning to the proof, we first formulate the $\mathbb{Z}_N$ analogues of Definition \ref{def:z2-em-exchanging-symmetry} and Lemma \ref{lem:local-stabilizer-basis-change}.

\begin{definition}[$e$-$m$ exchanging symmetry on the $\mathbb{Z}_N$ toric code]
\label{def:ZN-em-exchanging-symmetry}
We use the vertex--plaquette pairing $p(v)$ introduced in Section~2.2 to identify vertex and plaquette indices.
For finitely supported exponent column vectors $a$ and $b$ over $\mathbb{Z}_N$, define
$\mathsf{A}(a):=\prod_v A_v^{(a_v)}$ and
$\mathsf{B}(b):=\prod_p B_p^{(-b_p)}$.
The minus sign in the definition of $\mathsf{B}(b)$ is chosen so that the diagonal $e$-$m$ exchange corresponds to the identity map on the exponent labels.

A locality-preserving Clifford unitary $U$ is called an $e$-$m$ exchanging symmetry if there exist locality-preserving, locally invertible $\mathbb{Z}_N$-linear maps $F$ and $G$ such that
\begin{align}
U\mathsf{A}(a)U^\dagger=\mathsf{B}(Fa),
\qquad
U\mathsf{B}(b)U^\dagger=\mathsf{A}(Gb),
\label{eq:ZN-general-stabilizer-exchange}
\end{align}
for all $a$ and $b$, with $F\mathbf{1}=G\mathbf{1}=\mathbf{1}$.
Here, local invertibility means that $F^{-1}$ and $G^{-1}$ are also locality preserving.
\end{definition}

The local basis change lemma (Lemma \ref{lem:local-stabilizer-basis-change}) from Section~\ref{sec:z2-toric-code} has the following generalization to $\mathbb{Z}_N$.

\begin{lemma}[Local stabilizer basis change over $\mathbb{Z}_N$]
\label{lem:ZN-local-stabilizer-basis-change}
Let $F$ be a locality-preserving and locally invertible $\mathbb{Z}_N$-linear transformation of the oriented plaquette stabilizer generators satisfying $F\mathbf{1}=\mathbf{1}$.
Then there exists a locality-preserving CSS Clifford unitary $V_F$ such that
\begin{align}
V_FA_v^{(1)}V_F^\dagger=A_v^{(1)},
\qquad
V_F\mathsf{B}(b)V_F^\dagger=\mathsf{B}(Fb).
\label{eq:ZN-local-stabilizer-basis-change}
\end{align}
\end{lemma}

The proof is given in Appendix \ref{append:ZN-local-stabilizer-basis-change}.

\begin{proof}[Proof of Theorem \ref{thm:even-N-em-order-restated}]
Suppose, for contradiction, that $U_{em}^2\propto I$.
We first use Lemma \ref{lem:ZN-local-stabilizer-basis-change} to reduce the stabilizer action to an exact exchange of diagonal pairs.
With $F$ as in Eq.\,\eqref{eq:ZN-general-stabilizer-exchange}, define
\begin{align}
\widehat{U}:=V_F^\dagger U_{em}V_F.
\end{align}
Since $V_F$ fixes the vertex stabilizers and sends $\mathsf{B}(b)$ to $\mathsf{B}(Fb)$, we obtain
$\widehat{U}\mathsf{A}(a)\widehat{U}^\dagger=\mathsf{B}(a)$.
Applying $\widehat{U}$ once more and using $\widehat{U}^2\propto I$ gives the reverse exchange.
Thus,
\begin{align}
\widehat{U}A_v^{(1)}\widehat{U}^\dagger
=B_{p(v)}^{(-1)},
\qquad
\widehat{U}B_{p(v)}^{(-1)}\widehat{U}^\dagger
=A_v^{(1)}.
\label{eq:ZN-normalized-stabilizer-exchange}
\end{align}
After absorbing an overall phase, we may assume $\widehat{U}^2=I$.

We use the same L-shaped string arrangement as in Fig.\,\ref{fig:no-go-string-geometry}, with separations much larger than the locality range of $\widehat{U}$.
Orient $\alpha_e$ from $v_1$ to $v_2$ and $\beta_e$ from $v_2$ to $v_3$, with $e^{-1}$ at each initial endpoint and $e$ at each final endpoint.
Choose $\alpha_m$ and $\beta_m$ on the nearby dual paths with the same endpoint syndromes as $\widehat{U}\alpha_e\widehat{U}^\dagger$ and $\widehat{U}\beta_e\widehat{U}^\dagger$, respectively.
Since $\alpha_m$ and $\beta_e$ cross once, whereas $\alpha_e$ and $\beta_m$ are disjoint,
\begin{align}
\langle\alpha_m,\beta_e\rangle=\pm1,
\qquad
\langle\alpha_e,\beta_m\rangle=0.
\label{eq:ZN-alpha-m-beta-e}
\end{align}

The local stabilizer decomposition in Lemma \ref{lem:diagonal-exchange-string-dressing} extends to $\mathbb{Z}_N$ by the same contractible-region argument.
Separating the factors at the middle pair $h_2=(v_2,p_2)$ from the remaining stabilizers, we obtain
\begin{align}
\widehat{U}\alpha_e\widehat{U}^\dagger
&\propto\alpha_m A_{v_2}^{(x)}B_{p_2}^{(-y)}
\widetilde{S}_{\alpha_e},
\nonumber\\
\widehat{U}\alpha_m\widehat{U}^\dagger
&\propto\alpha_e A_{v_2}^{(-y)}B_{p_2}^{(x)}
\widetilde{S}_{\alpha_m},
\nonumber\\
\widehat{U}\beta_e\widehat{U}^\dagger
&\propto\beta_m A_{v_2}^{(z)}B_{p_2}^{(-w)}
\widetilde{S}_{\beta_e},
\qquad x,y,z,w\in\mathbb{Z}_N.
\label{eq:ZN-string-images}
\end{align}
The second relation follows by applying $\widehat{U}$ once more to the first and using $\widehat{U}^2=I$ together with Eq.\,\eqref{eq:ZN-normalized-stabilizer-exchange}.
Each tilded operator is a product of stabilizers near its corresponding string and contains neither $A_{v_2}^{(1)}$ nor $B_{p_2}^{(1)}$.

As in the $\mathbb{Z}_2$ proof, preservation of the commutation exponent gives
\begin{align}
\langle\alpha_m,\beta_e\rangle
&=
\langle\widehat{U}\alpha_m\widehat{U}^\dagger,
\widehat{U}\beta_e\widehat{U}^\dagger\rangle
\nonumber\\
&=
z\langle\alpha_e,A_{v_2}^{(1)}\rangle
+x\langle B_{p_2}^{(1)},\beta_m\rangle.
\end{align}
The tilded operators make no contribution because they exclude the middle generators and are supported far from the remaining endpoints.
Both endpoint commutation exponents on the right are $\pm1$, since the endpoints have unit charge.
For even $N$, reduction modulo two is well defined and removes these orientation signs.
Together with Eq.\,\eqref{eq:ZN-alpha-m-beta-e}, this gives
\begin{align}
x+z=1\pmod{2}.
\label{eq:ZN-commutation-condition}
\end{align}

We now derive another constraint by closing the strings into contractible loops.
Choose the oriented closing strings $\gamma_e$ and $\gamma_m$ as in Fig.\,\ref{fig:no-go-string-geometry}(b,c), sufficiently far from $h_2$ that
$\widehat{U}\gamma_e\widehat{U}^\dagger\propto\gamma_m\widetilde{S}_{\gamma_e}$,
where $\widetilde{S}_{\gamma_e}$ contains neither middle generator.
The electric loop is a product of oriented plaquette stabilizers in $M$, with $p_2\notin M$.
Its image is therefore a product of vertex stabilizers with no factor at $v_2$, since each plaquette maps to the vertex immediately to its lower left.
The bare magnetic loop also encloses no $v_2$.

We can now compute the transformed electric loop in a second way, by multiplying the three transformed segments.
The first and third relations in Eq.\,\eqref{eq:ZN-string-images} give a factor $A_{v_2}^{(x+z)}$, while neither the bare magnetic loop nor the tilded products contributes at $v_2$.
All decompositions are taken in a common contractible region smaller than the torus, so global redundancy cannot change a local exponent.
Comparing the exponent at $v_2$ in the two disk-supported decompositions yields
\begin{align}
x+z=0\pmod{N}.
\label{eq:ZN-loop-condition}
\end{align}
Because $N$ is even, this contradicts Eq.\,\eqref{eq:ZN-commutation-condition}.
Hence $U_{em}^2\not\propto I$.
\end{proof}

\subsection{Order-four Clifford realization for even $N$}

The no-go theorem above shows that the order of any Clifford realization of the $e$-$m$ exchange must be at least four for even $N$.
This lower bound is tight.
An explicit order-four Clifford realization is constructed in Appendix \ref{app:order-four-exchange}.

\begin{figure}
\centering
\raisebox{-54.45436pt}[525.03552pt][54.45436pt]{%
\includegraphics[width=491.43622pt,height=579.48988pt]{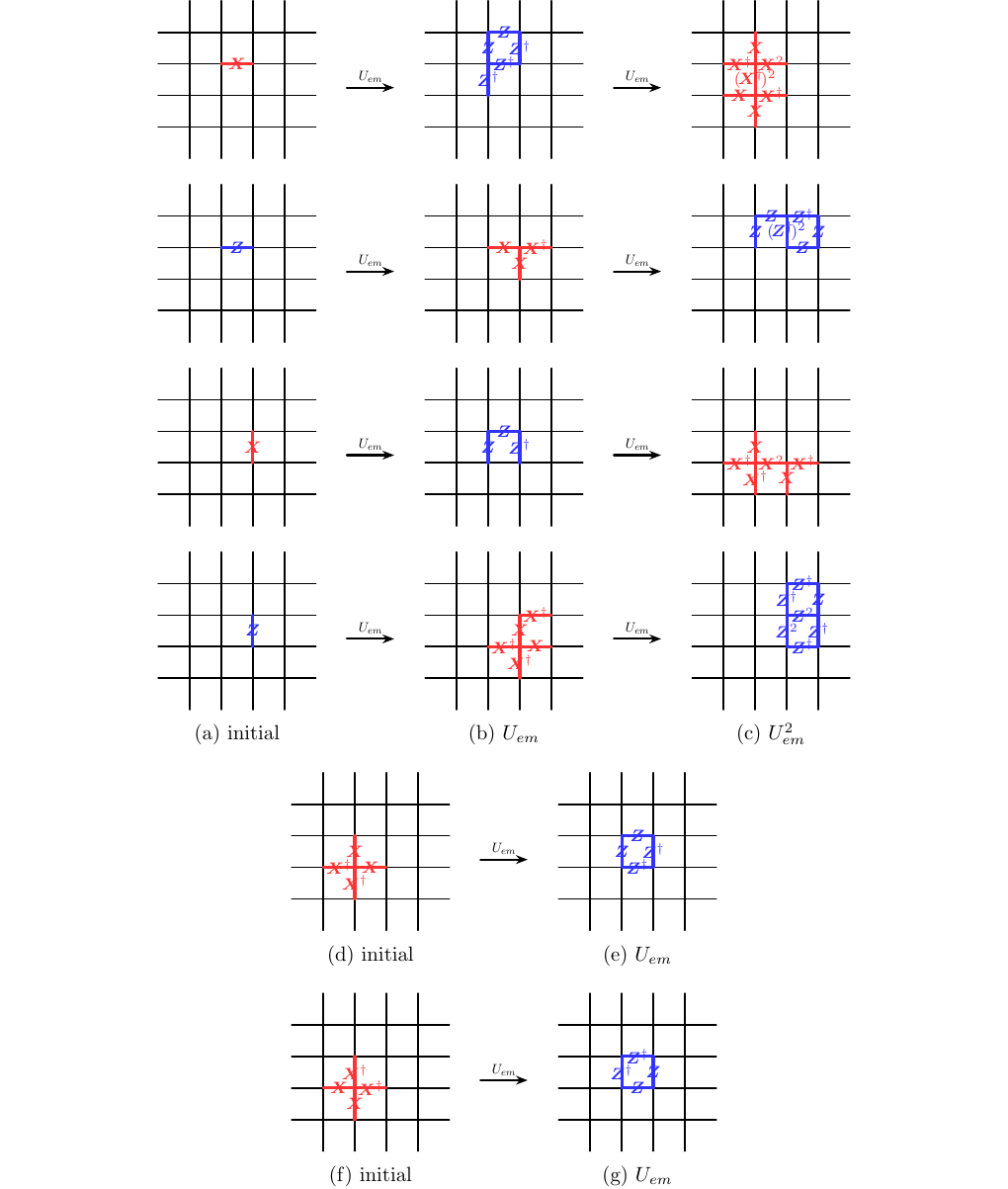}}
\caption{Conjugation rules for the explicit $\mathbb{Z}_N$ exchange $U_{em}$.
Panels (a)--(c) show the first two iterates of representative local Pauli operators, while panels (d)--(g) show the exchange of the two oriented vertex operators with their paired plaquette operators.
Red and blue labels denote $X$- and $Z$-type factors, respectively, and a dagger denotes the inverse generalized Pauli operator.}
\label{fig:u_action_qd}
\end{figure}

\subsection{Non-Clifford $\mathbb{Z}_2$ realization}

Despite the Clifford obstruction, the exchange $U_{em}$ in
Eq.\,\eqref{eq:zn-toric-code-em-exchange} can be converted into an exact
$\mathbb{Z}_2$ symmetry for every $N$ by a non-Clifford dressing.
Define
\begin{align}
\widetilde{U}_{em}:=U_{em}D_N,
\qquad
D_N:=\prod_{v\in V(\Lambda)}
\sum_{a,b\in\mathbb{Z}_N}
\phi(a,b)\,\Pi_v(a)\Pi_{p(v)}(b),
\end{align}
where
\begin{gather}
\phi(a,b):=e^{\frac{\pi i[-ab]}{N}},
\qquad
\Pi_v(a):=\frac{1}{N}\sum_{r=0}^{N-1}
\omega^{-ar}A_v^{(r)},
\qquad
\Pi_p(b):=\frac{1}{N}\sum_{s=0}^{N-1}
\omega^{-bs}B_p^{(s)}.
\end{gather}
Here, $[k]\in\{0,\ldots,N-1\}$ denotes the standard representative
of $k\in\mathbb{Z}_N$.
This choice of $\phi$ satisfies
\begin{align}
\phi(a,b)\phi(-b,-a)=\omega^{-ab}.
\end{align}
For $N=2$, the dressing operator $D_N$ reduces to $D$ in
Eq.\,\eqref{eq:z2-non-clifford-dressing}.

\begin{proposition}
The dressed exchange satisfies
\begin{align}
\widetilde{U}_{em}^{\,2}\propto I.
\end{align}
\end{proposition}

\begin{proof}
It follows from Eq.\,\eqref{eq:anyon_permutation_sym} that
\begin{align}
U_{em}D_NU_{em}^\dagger
=
\prod_v\sum_{a,b}
\phi(-b,-a)\Pi_v(a)\Pi_{p(v)}(b).
\end{align}
Therefore,
\begin{align}
\bigl(U_{em}D_NU_{em}^\dagger\bigr)D_N=K,
\end{align}
where
\begin{align}
K:=
\prod_v\sum_{a,b}
\omega^{-ab}\Pi_v(a)\Pi_{p(v)}(b).
\end{align}
Since $U_{em}^2$ commutes with $D_N$, we obtain
\begin{align}
\widetilde{U}_{em}^{\,2}
&=
U_{em}D_NU_{em}D_N
\nonumber\\
&=
\bigl(U_{em}D_NU_{em}^\dagger\bigr)U_{em}^2D_N
\nonumber\\
&=
\bigl(U_{em}D_NU_{em}^\dagger\bigr)D_NU_{em}^2
=
KU_{em}^2.
\end{align}
The operator $K$ cancels the stabilizer dressing produced by
$U_{em}^2$.  For example,
\begin{align}
KX_{h_{0,0}}K^\dagger
=
X_{h_{0,0}}
A_{(0,0)}^{(-1)}A_{(0,-1)}^{(1)}.
\end{align}
The same cancellation holds for the remaining single-edge Pauli
generators $Z_h$, $X_v$, and $Z_v$.
Consequently, $KU_{em}^2$ commutes with every Pauli operator and is therefore proportional to the identity.
Hence,
\begin{align}
\widetilde{U}_{em}^{\,2}\propto I.
\end{align}
\end{proof}

\section{Outlook}

In this paper, we proved a no-go theorem for order-two Clifford realizations of electric--magnetic duality in the toric code.

\paragraph{$e$-$m$ exchange on general lattices.}

We have focused on the toric code on a square lattice. 
Our definition of microscopic $e$-$m$ exchange assumes a bijection between the vertex and plaquette stabilizer groups, making the square lattice a particularly natural setting.
It would nevertheless be interesting to formulate an analogous notion on more general lattices.
For example, on a closed honeycomb lattice, the vertex and plaquette stabilizer groups have different numbers of independent generators, so no unitary can exchange the two groups bijectively in the sense of Definition~\ref{def:z2-em-exchanging-symmetry}.

This counting obstruction does not by itself rule out a microscopic realization of the anyon permutation.
Indeed, one may relate toric-code realizations on different lattices after adding ancillary qubits and applying local transformations.
From this perspective, the square-lattice $e$-$m$ exchange should have an analogue on the honeycomb lattice once appropriate auxiliary degrees of freedom are included.
A broader microscopic definition of $e$-$m$ exchange may therefore need to allow ancillary qubits, or more generally finite-depth local quantum channels rather than unitary automorphisms of a fixed microscopic Hilbert space.

\paragraph{Odd--even effect.}

For the $\mathbb Z_N$ toric code, we found that the obstruction depends only on whether $N$ is odd or even.
This parity dependence is suggestive in view of the ground-state degeneracy on the real projective plane $\mathbb{RP}^2$.
The degeneracy is one for odd $N$ and two for even $N$.
Indeed, the ground-state degeneracy is closely related to the mod-two self-intersection structure of $\mathbb{RP}^2$, which can be characterized in terms of the Wu class.
In two dimensions, the relevant Wu class is the first Stiefel--Whitney class $w_1$, whose Poincar\'e dual detects the orientation-reversing cycle and its nontrivial self-intersection.

The similarity to our odd--even Clifford obstruction is tempting, although its precise meaning remains unclear.
The Wu-class and ground-state-degeneracy arguments are properties of the topological theory and do not by themselves immediately distinguish Clifford from non-Clifford microscopic realizations.
Moreover, the relation between the $e$-$m$ exchange defined microscopically on the square-lattice torus and an exchange operation defined on $\mathbb{RP}^2$ is not automatic.

Our proof suggests a somewhat different perspective.
The obstruction appears through the stabilizer decorations attached to the endpoints of electric and magnetic string operators.
The particular L-shaped string configuration used in the proof probes these decorations through a forced crossing, locally reproducing a geometric feature reminiscent of a cross-cap without ever changing the topology of the underlying lattice.
Understanding whether this local obstruction has a direct field-theoretic interpretation in terms of nonorientable geometry or Wu classes would be an interesting direction for future work.

\paragraph{Beyond electric--magnetic exchange.}

It would also be interesting to understand how general the order obstruction is among anyon-permuting symmetries of topological stabilizer codes.
The proof presented here relies primarily on the endpoint structure and mutual braiding of electric and magnetic string operators, suggesting that analogous obstructions may occur for other anyon permutations or in more general quantum double and CSS-type models.
A natural question is whether the braiding data of the emergent anyon theory, together with the microscopic Pauli structure, determine when an anyon permutation admits a faithful Clifford realization and when its order must increase microscopically.

\paragraph{Non-Clifford resources and magic.}

The Clifford assumption is essential to our obstruction.
Once non-Clifford operations are allowed, the electric--magnetic exchange can be dressed into an exact order-two symmetry for arbitrary $N$.
The required dressing is nevertheless highly structured, depending only on the local electric and magnetic stabilizer eigenvalues.
This raises the question of how much non-Clifford resource is actually required to realize the emergent symmetry faithfully.
For example, one may ask whether the minimal Clifford-hierarchy level or amount of magic required to remove the obstruction can be characterized systematically.
It would be particularly interesting to understand whether such a resource cost is determined by the braiding data of the emergent anyons or by finer details of the microscopic realization.

\section*{Acknowledgement}
We thank Ryohei Kobayashi for useful discussions.
Research at Perimeter Institute is supported in part by the Government of Canada through the Department of Innovation, Science and Economic Development and by the Province of Ontario through the Ministry of Colleges and Universities.
This work was supported by the RIKEN TRIP initiative and JSPS KAKENHI Grant Number 23KJ1154, 24K17047.

\appendix

\section{Local Pauli operators under $U_{em}^2$} 
\label{app:2vs4}

We characterize the local Pauli operators fixed, up to a phase, by two
conjugations with the explicit exchange $U_{em}$.
The argument applies to every $N\geq 2$, with the $N=2$ case reproducing
the qubit exchange of Section 2.
Throughout this appendix, the support of a local operator and its
fixed-width neighborhood are assumed to lie within a disk-shaped region
smaller than the torus.

\begin{proposition}
Let $U_{em}$ be the explicit Clifford exchange in
Eq.\,\eqref{eq:zn-toric-code-em-exchange}, and let $P$ be a local Pauli
operator.
Then
\begin{align}
U_{em}^2PU_{em}^{-2}\propto P
\end{align}
if and only if $P$ is proportional to a product of the toric-code
stabilizer generators $A_v^{(1)}$ and $B_p^{(1)}$.
\end{proposition}

\begin{proof}
For each vertex--plaquette pair $(v,p(v))$, define the commutation
exponents
\begin{align}
s_v:=\langle A_v^{(1)},P\rangle,
\qquad
t_v:=\langle B_{p(v)}^{(1)},P\rangle.
\end{align}
Multiplying the single-edge conjugation rules in
Fig.\,\ref{fig:u_action_qd} gives
\begin{align}
U_{em}^2PU_{em}^{-2}
\propto
P\prod_v
\bigl(A_v^{(1)}\bigr)^{t_v}
\bigl(B_{p(v)}^{(1)}\bigr)^{s_v}.
\label{eq:local-pauli-square-dressing}
\end{align}
Thus, the stabilizer dressing is determined by the local syndrome of $P$.

Suppose that $U_{em}^2PU_{em}^{-2}\propto P$.
Equation \eqref{eq:local-pauli-square-dressing} then implies that the
stabilizer product on the right is proportional to the identity.
Its $X$- and $Z$-type parts must separately be trivial.
Since $s_v$ and $t_v$ are supported near $P$, neither part can involve
a nontrivial global redundancy.
Consequently,
\begin{align}
s_v=t_v=0
\qquad\text{for every }v.
\end{align}
Hence $P$ commutes with every vertex and plaquette stabilizer.
By the same local stabilizer-decomposition argument used in
Lemma \ref{lem:diagonal-exchange-string-dressing}, now over
$\mathbb{Z}_N$, $P$ is proportional to a product of stabilizers.

Conversely, $U_{em}^2$ fixes every stabilizer generator, so it fixes
every such product.
\end{proof}

For $N=2$, $U_{em}^4\propto I$.
Therefore every local Pauli operator not proportional to a stabilizer
product has projective period four under conjugation by $U_{em}$.

\section{Local stabilizer basis changes over $\mathbb{Z}_N$} \label{append:ZN-local-stabilizer-basis-change}
This appendix proves Lemma \ref{lem:ZN-local-stabilizer-basis-change} from Section~\ref{sec:zn-toric-code}.
We first introduce the $\mathbb{Z}_N$ analogue of the binary symplectic representation of Pauli operators and then use this formalism to construct the required local Clifford basis change.

\subsection{Pauli symplectic formalism over $\mathbb{Z}_N$} \label{subappend:zn-pauli-symplectic-formalism}
Let $\omega=e^{2\pi i/N}$ and consider $n$ qudits with computational basis $\{|q\rangle:q\in\mathbb{Z}_N^n\}$.
For $z,x\in\mathbb{Z}_N^n$, define
\begin{align}
Z(z):=\prod_{j=1}^n Z_j^{z_j},
\qquad
X(x):=\prod_{j=1}^n X_j^{x_j},
\qquad
P(z,x):=Z(z)X(x).
\label{eq:ZN-Pauli-labels}
\end{align}
The relation $ZX=\omega XZ$ implies
\begin{align}
P(z,x)P(z',x')
=\omega^{z^tx'-x^tz'}P(z',x')P(z,x).
\label{eq:ZN-Pauli-commutator}
\end{align}
Thus Pauli operators modulo phase are labeled by $u=(z,x)^t\in\mathbb{Z}_N^{2n}$, with commutation pairing
\begin{align}
\langle u,u'\rangle
:=u^tJu',
\qquad
J:=
\begin{pmatrix}
0&I\\
-I&0
\end{pmatrix}.
\label{eq:ZN-symplectic-form}
\end{align}
A Clifford unitary transforms these labels by an invertible linear map $S$ that preserves the commutation pairing in Eq.\,\eqref{eq:ZN-Pauli-commutator}, so
\begin{align} \label{eq:ZN-symplectic-condition}
S^tJS=J\mod{N}.
\end{align}
We only need the CSS transformation associated with an invertible matrix $T\in\operatorname{GL}(n,\mathbb{Z}_N)$,
\begin{align}
S_T:=
\begin{pmatrix}
T&0\\
0&(T^{-1})^t
\end{pmatrix}.
\label{eq:ZN-CSS-symplectic-map}
\end{align}
Its symplecticity follows from
\begin{align}
S_T^tJS_T
=
\begin{pmatrix}
T^t&0\\
0&T^{-1}
\end{pmatrix}
\begin{pmatrix}
0&I\\
-I&0
\end{pmatrix}
\begin{pmatrix}
T&0\\
0&(T^{-1})^t
\end{pmatrix}
=J.
\end{align}
In this CSS case the corresponding Clifford unitary has the explicit action
\begin{align}
V_T|q\rangle
:=|(T^{-1})^tq\rangle,
\qquad q\in\mathbb{Z}_N^n.
\label{eq:ZN-CSS-basis-permutation}
\end{align}
The Pauli generators transform under $V_T$ as
\begin{align}
V_TZ(z)V_T^\dagger=Z(Tz),
\qquad
V_TX(x)V_T^\dagger=X((T^{-1})^tx).
\label{eq:ZN-CSS-Pauli-action}
\end{align}
If $T$ and $T^{-1}$ are locality preserving, Eq.\,\eqref{eq:ZN-CSS-Pauli-action} shows that $V_T$ is locality preserving as well.

\subsection{Proof of the local basis change lemma}
We restate Lemma \ref{lem:ZN-local-stabilizer-basis-change} so that its assumptions and conclusion are available alongside the proof.
\begingroup
\renewcommand{\thelemma}{\getrefnumber{lem:ZN-local-stabilizer-basis-change}}
\renewcommand{\theHlemma}{restated.lem:ZN-local-stabilizer-basis-change}
\begin{lemma}[Restated] \label{lem:ZN-local-stabilizer-basis-change-restated}
Let $F$ be a locality-preserving and locally invertible $\mathbb{Z}_N$-linear transformation of the oriented plaquette stabilizer generators satisfying $F\mathbf{1}=\mathbf{1}$.
Then there is a locality-preserving CSS Clifford unitary $V_F$ such that
\begin{align}
V_FA_v^{(1)}V_F^\dagger=A_v^{(1)},
\qquad
V_F\mathsf{B}(b)V_F^\dagger=\mathsf{B}(Fb).
\end{align}
\end{lemma}
\addtocounter{lemma}{-1}
\endgroup
\begin{proof}[Proof of Lemma \ref{lem:ZN-local-stabilizer-basis-change-restated}]
We follow the construction of Lemma \ref{lem:local-stabilizer-basis-change}, retaining the oriented incidence numbers that disappear over $\mathbb{F}_2$.
All modules in this proof have coefficients in $\mathbb{Z}_N$.
Let $C_0:=\bigoplus_{v\in V(\Lambda)}\mathbb{Z}_Nv$, $C_1:=\bigoplus_{e\in E(\Lambda)}\mathbb{Z}_Ne$, $C_2:=\bigoplus_{p\in F(\Lambda)}\mathbb{Z}_Np$ be the vertex, edge, and plaquette modules, and let
\begin{align}
G_1:C_0\longrightarrow C_1,
\qquad
G_2:C_2\longrightarrow C_1
\end{align}
be the oriented vertex and plaquette generator matrices.
Write $G_2(e,p)\in\{0,\pm1\}$ for the coefficient of the oriented edge $e$ in the boundary of $p$.
The all-ones plaquette vector is
\begin{align}
\mathbf{1}:=\sum_{p\in F(\Lambda)}p
\in C_2\cong\mathbb{Z}_N^{|F(\Lambda)|\times1}.
\label{eq:ZN-all-ones-plaquette-vector}
\end{align}
Set $L:=F-I$.
The condition $F\mathbf{1}=\mathbf{1}$ implies
\begin{align}
L\mathbf{1}=0.
\label{eq:ZN-L-augmentation}
\end{align}
Let $\ell_p\in C_2^*$ be the $p$-th row of $L$.
Eq.\,\eqref{eq:ZN-L-augmentation} states that
\begin{align}
\sum_q\ell_p(q)=\ell_p\mathbf{1}=0
\qquad\text{in }\mathbb{Z}_N.
\label{eq:ZN-row-coefficient-sum}
\end{align}
Let $R$ be the locality range of $F$, and let $P_R(p)$ be the set of plaquettes reachable from $p$ by at most $R$ edge-sharing steps.
Then $\operatorname{supp}(\ell_p)\subset P_R(p)$.
For each $q\in P_R(p)\setminus\{p\}$, choose a simple path of plaquettes
\begin{align}
p=p_0,p_1,\ldots,p_k=q,
\qquad
k\leq R,
\end{align}
and denote the edge shared by $p_{j-1}$ and $p_j$ by $e_j$.
Let $e^*\in C_1^*$ denote the row vector dual to the edge basis element $e\in C_1$.
Similarly, let $p^*,q^*\in C_2^*$ denote the row vectors dual to the plaquette basis elements $p,q\in C_2$.
Define
\begin{align}
s_q:=\sum_{j=1}^kG_2(e_j,p_j)e_j^*
\in C_1^*.
\label{eq:ZN-signed-path-vector}
\end{align}
The two plaquettes incident to $e_j$ have opposite incidence numbers.
\begin{align}
G_2(e_j,p_{j-1})=-G_2(e_j,p_j).
\label{eq:ZN-opposite-incidence}
\end{align}
At an intermediate plaquette $p_j$, the two contributions to $s_qG_2$ cancel because
\begin{align}
G_2(e_j,p_j)^2
+G_2(e_{j+1},p_{j+1})G_2(e_{j+1},p_j)
=1-1=0.
\end{align}
The coefficients at $p_0=p$ and $p_k=q$ are $-1$ and $+1$, respectively, while every plaquette outside the path has coefficient zero.
Hence
\begin{align}
s_qG_2=q^*-p^*.
\label{eq:ZN-path-boundary}
\end{align}
Eq.\,\eqref{eq:ZN-row-coefficient-sum} permits the decomposition
\begin{align}
\ell_p
=\sum_{q\in P_R(p)\setminus\{p\}}\ell_p(q)(q^*-p^*).
\end{align}
Set
\begin{align}
h_p:=\sum_{q\in P_R(p)\setminus\{p\}}\ell_p(q)s_q.
\end{align}
Eq.\,\eqref{eq:ZN-path-boundary} implies
\begin{align}
h_pG_2
=\sum_{q\in P_R(p)\setminus\{p\}}\ell_p(q)(q^*-p^*)
=\ell_p.
\end{align}
Every edge in $\operatorname{supp}(h_p)$ lies within distance $R$ of $p$.
The vectors $h_p$ can therefore be assembled as the rows of a locality-preserving map $H:C_1\to C_2$, and they obey
\begin{align}
L=HG_2.
\label{eq:ZN-L-HG2}
\end{align}

Define the transformation of edge labels
\begin{align}
T:=I+G_2H.
\end{align}
Eq.\,\eqref{eq:ZN-L-HG2} implies
\begin{align}
TG_2
=G_2+G_2HG_2
=G_2(I+L)
=G_2F.
\end{align}
The CSS relation $G_2^tG_1=0$ also implies
\begin{align}
T^tG_1
=(I+H^tG_2^t)G_1
=G_1.
\end{align}
Since $T$ is invertible, the last identity is equivalent to
\begin{align}
(T^{-1})^tG_1=G_1.
\label{eq:ZN-T-relations}
\end{align}
The inverse contains the minus sign that is invisible over $\mathbb{F}_2$.
\begin{align}
T^{-1}=I-G_2F^{-1}H.
\label{eq:ZN-T-inverse}
\end{align}
To verify this identity, use $HG_2=F-I$ and compute
\begin{align}
(I+G_2H)(I-G_2F^{-1}H)
&=I+G_2\bigl(H-F^{-1}H-HG_2F^{-1}H\bigr)
\nonumber\\
&=I+G_2\bigl(H-F^{-1}H-(F-I)F^{-1}H\bigr)
=I.
\end{align}
Thus $T$ and $T^{-1}$ are both locality preserving.
Apply the CSS Clifford unitary $V_T$ from Eq.\,\eqref{eq:ZN-CSS-basis-permutation} to the edge qudits and denote it by $V_F$. Eqs.\,\eqref{eq:ZN-CSS-Pauli-action} and \eqref{eq:ZN-T-relations} show that $V_F$ fixes every oriented vertex generator and transforms the plaquette module by $F$, as required in Eq.\,\eqref{eq:ZN-local-stabilizer-basis-change}.

\end{proof}

\section{Order-four Clifford exchange for even $N$}
\label{app:order-four-exchange}

For even $N$, Theorem~\ref{thm:even-N-em-order-restated} rules out an order-two Clifford realization of the $e$-$m$ exchange. 
Here, we construct an order-four realization. 
We define
\begin{align}
W:=R\,U_{em}\,D,
\label{eq:rd-overview-W}
\end{align}
where $U_{em}$ is the order-$2N$ exchange in Eq.~\eqref{eq:zn-toric-code-em-exchange}, and $R$ and $D$ are locality-preserving Clifford dressings defined below.

\begin{figure}[!t]
\centering
\makebox[\textwidth][c]{%
\includegraphics[scale=1]{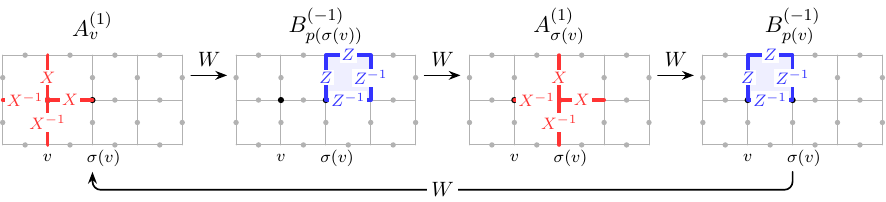}}
\caption{Transformation of a vertex stabilizer $A_v^{(1)}$ under $W$, with $v=(2i,j)$ and
$\sigma(v)=(2i+1,j)$. After two applications it becomes
$A_{\sigma(v)}^{(1)}$, and after four it returns to $A_v^{(1)}$.}
\label{fig:rd-W-stabilizer-orbit}
\end{figure}

We work on a finite torus with even $L_x\geq 2$ and arbitrary $L_y\geq 2$.
Let $\sigma$ be the exchange of the neighboring vertices.
\[
(2i,j)\longleftrightarrow(2i+1,j)
\]
for every $i,j$.
The dressings are chosen so that
\begin{align}
W A_v^{(1)}W^{\dagger}
=B_{p(\sigma(v))}^{(-1)},\qquad
W B_{p(v)}^{(-1)}W^{\dagger}
=A_v^{(1)}.
\label{eq:rd-overview-W-exchange}
\end{align}
Thus, the action of $W$ on the stabilizer generators has order four, as illustrated in Fig.~\ref{fig:rd-W-stabilizer-orbit}.

The two dressings have different roles.
The Clifford unitary $R$ fixes all vertex operators and exchanges the adjacent plaquette operators as specified by the exchange $\sigma$.
The Clifford unitary $D$ fixes every vertex and plaquette operator, but changes the action on Pauli operators outside the stabilizer group.
It is designed specifically to cancel the residual action of $(R U_{em})^4$.

\subsection{Clifford dressings}
\label{subapp:rd-symplectic-matrices}

We use the Pauli symplectic notation of Appendix~\ref{subappend:zn-pauli-symplectic-formalism}.
Let $a_v,b_p\in\mathbb{Z}_N^{|E(\Lambda)|}$ denote the edge-label vectors
\begin{align}
A_v^{(1)}=X(a_v),\qquad B_p^{(1)}=Z(b_p).
\end{align}
They satisfy $b_p^t a_v=0$ for every $p,v$.
Let $\mathbf e_{i,j}$ be the basis vector associated with the vertical edge shared by $p(2i,j)$ and $p(2i+1,j)$.

For a computational-basis state
$|q\rangle=\bigotimes_{e\in E(\Lambda)}|q_e\rangle_e$,
define
\begin{align}
D|q\rangle=|T_Dq\rangle, \qquad
R|q\rangle=|T_Rq\rangle,
\label{eq:rd-overview-basis-action}
\end{align}
with
\begin{align}
T_D&=I-\sum_{i,j}a_{(2i,j)}b_{p(2i,j)}^t,\nonumber\\
T_R&=I-\sum_{i,j}\mathbf e_{i,j}
\bigl(b_{p(2i,j)}-b_{p(2i+1,j)}\bigr)^t.
\label{eq:rd-overview-computational-matrices}
\end{align}
The inverses are
\begin{align}
T_D^{-1}
=I+\sum_{i,j}a_{(2i,j)}b_{p(2i,j)}^t,
\qquad
T_R^{-1}&=T_R.
\end{align}
As reviewed in Appendix~\ref{subappend:zn-pauli-symplectic-formalism}, an invertible edge transformation $T$ induces the CSS Clifford action
\[
z\mapsto (T^{-1})^t z,\qquad x\mapsto Tx.
\]
Since $T_D,T_R$ and their inverses are local, $D$ and $R$ are locality-preserving Clifford unitaries.

The definitions immediately give
\[
D A_v^{(1)}D^\dagger=A_v^{(1)},\qquad
D B_p^{(1)}D^\dagger=B_p^{(1)},
\]
while $R$ fixes every vertex operator and exchanges
\[
B_{p(v)}^{(1)}
\longleftrightarrow
B_{p(\sigma(v))}^{(1)}.
\]
Combining these relations with the stabilizer action of $U_{em}$ gives Eq.~\eqref{eq:rd-overview-W-exchange}.
In particular,
\begin{align}
W^2A_v^{(1)}W^{-2}
=A_{\sigma(v)}^{(1)},\qquad
W^2B_{p(v)}^{(1)}W^{-2}
=B_{p(\sigma(v))}^{(1)}.
\label{eq:rd-overview-round-trip}
\end{align}
Since $\sigma^2=I$, the stabilizer action has order four.

\subsection{Fourth power on full Pauli algebra}

It remains to show that the full microscopic action also has order four.
Set
\[
V:=R U_{em},
\qquad
W=VD.
\]
Because $D$ fixes every stabilizer, $V$ and $W$ have the same action on stabilizers.
Thus, after four applications of $V$, every stabilizer returns to itself, but a generic Pauli operator can still acquire a stabilizer dressing.

Using the single-edge conjugation rules under $U_{em}$ in Fig.~\ref{fig:u_action_qd} together with the action of $R$, direct iteration gives, for example,
\begin{align}
V^4X_{h_{0,0}}V^{-4}
=
X_{h_{0,0}}A_{(0,0)}^{(1)}A_{(1,0)}^{(1)}
\bigl(A_{(0,-1)}^{(1)}A_{(1,-1)}^{(1)}\bigr)^{-1}.
\label{eq:rd-bare-fourth-edge-example}
\end{align}
The analogous calculations for the elementary horizontal and vertical \(X\)- and \(Z\)-operators can be summarized by introducing
\begin{align}
M:=
\sum_{i,j}
\bigl(a_{(2i,j)}+a_{(2i+1,j)}\bigr)
\bigl(b_{p(2i,j)}+b_{p(2i+1,j)}\bigr)^t.
\label{eq:rd-residual-matrix}
\end{align}
Since $b_p^t a_v=0$ for all $p,v$, every product of two summands vanishes and hence
\[ 
M^2=0.
\]
For arbitrary Pauli exponent vectors,
\begin{align}
V^4X(x)V^{-4}
=X\bigl((I+M)x\bigr),\qquad
V^4Z(z)V^{-4}
=Z\bigl((I-M^t)z\bigr).
\label{eq:rd-bare-fourth-power}
\end{align}
Equivalently, on the computational basis,
\begin{align}
V^4|q\rangle=e^{i\theta}|(I+M)q\rangle
\label{eq:rd-bare-fourth-basis}
\end{align}
for an overall phase $e^{i\theta}$ independent of $q$.

We now show that the four insertions of $D$ cancel this residual transformation.
Consider a simultaneous stabilizer eigenspace,
\begin{align}
A_v^{(1)}|\chi\rangle
=\omega^{\alpha_v}|\chi\rangle,
\qquad
B_{p(v)}^{(1)}|\chi\rangle
=\omega^{\beta_v}|\chi\rangle,
\qquad
\alpha_v,\beta_v\in\mathbb Z_N.
\end{align}
One can verify that, within this eigenspace, $D$ acts diagonally as
\begin{align}
D|\chi\rangle
=
\omega^{-\sum_{v=(2i,j)}\alpha_v\beta_v}
|\chi\rangle.
\label{eq:D-stabilizer-eigenvalue}
\end{align}
Let $w=\sigma(v)$. Since $V=R U_{em}$ permutes the stabilizer
generators,
\[
A_v^{(1)}\longmapsto
\bigl(B_{p(w)}^{(1)}\bigr)^{-1},
\qquad
B_{p(v)}^{(1)}\longmapsto
\bigl(A_v^{(1)}\bigr)^{-1},
\]
each conjugate $V^\ell D V^{-\ell}$ is again diagonal in the
simultaneous stabilizer eigenspaces.
Using Eq.~\eqref{eq:D-stabilizer-eigenvalue}, the four conjugates
give, for the pair $(v,w)$, the phases
\[
\omega^{-\alpha_v\beta_v},\qquad
\omega^{-\alpha_v\beta_w},\qquad
\omega^{-\alpha_w\beta_w},\qquad
\omega^{-\alpha_w\beta_v}.
\]
Hence
\begin{align}
\left[
\prod_{\ell=0}^{3}
V^{\ell} D V^{-\ell}
\right]
|\chi\rangle
=
\omega^{-\sum_{v=(2i,j)}
(\alpha_v+\alpha_{\sigma(v)})
(\beta_v+\beta_{\sigma(v)})}
|\chi\rangle .
\label{eq:rd-four-dressings-eigenvalue}
\end{align}
Using
\[
A_v^{(1)}A_w^{(1)}=X(a_v+a_w),
\qquad
\beta_v+\beta_w
=
\bigl(b_{p(v)}+b_{p(w)}\bigr)^tq,
\]
the phase above is equivalently implemented, in the computational basis, by
\[
q\longmapsto
q-\sum_{v=(2i,j)}
(a_v+a_{\sigma(v)})
\bigl(b_{p(v)}+b_{p(\sigma(v))}\bigr)^tq
=(I-M)q.
\]
Therefore,
\begin{align}
\left[
\prod_{\ell=0}^{3}V^\ell D V^{-\ell}
\right]|q\rangle
=
|(I-M)q\rangle.
\label{eq:rd-four-dressings-basis}
\end{align}

Finally, since $V^4$ fixes every stabilizer, it also fixes $D$ under conjugation.
Moreover, the conjugates $V^\ell D V^{-\ell}$ are functions of the
commuting stabilizers and hence commute. Therefore,
\begin{align}
W^4=(VD)^4
=
\left[
\prod_{\ell=0}^{3}V^\ell D V^{-\ell}
\right]V^4.
\end{align}
Using Eqs.~\eqref{eq:rd-four-dressings-basis} and
\eqref{eq:rd-bare-fourth-basis},
\begin{align}
W^4|q\rangle
&=
e^{i\theta}|(I-M)(I+M)q\rangle
=
e^{i\theta}|q\rangle,
\end{align}
where $M^2=0$.
Thus $W^4\propto I$.
Since $W^2$ exchanges the two vertex operators in each pair, $W$ has order four up to phase.

\mciteSetMidEndSepPunct{}{\ifmciteBstWouldAddEndPunct.\else\fi}{\relax}

\providecommand{\href}[2]{#2}\begingroup\raggedright\endgroup

%
\end{document}